\documentclass[11pt]{article}

\usepackage[margin=1in]{geometry}

\usepackage[T1]{fontenc}
\usepackage[utf8]{inputenc}
\usepackage{lmodern}
\usepackage{microtype}

\usepackage{amsmath, amsfonts, amssymb, amsthm, mathtools}
\theoremstyle{plain}
\newtheorem{assumption}{Assumption}
\newtheorem{theorem}{Theorem}

\newtheorem{result}{Result}

\usepackage{algorithm}
\usepackage{algpseudocode}

\usepackage{graphicx}
\usepackage{booktabs}
\usepackage{multicol}
\usepackage{multirow}
\graphicspath{{figures/}} 

\usepackage[hidelinks]{hyperref}
\usepackage[capitalize,nameinlink]{cleveref}
\usepackage{float}

\usepackage[authoryear]{natbib}

\title{Sequential Testing for Assessing the Incremental Value of Biomarkers Under Biorepository Specimen Constraints with Robustness to Model Misspecification}

\author{
  Indrila Ganguly\thanks{Corresponding author: \href{mailto:iganguly@fredhutch.org}{iganguly@fredhutch.org}}
  \and
  Ying Huang\thanks{\href{mailto:yhuang@fredhutch.org}{yhuang@fredhutch.org}}
}
\date{} 
\begin{document}

\maketitle
\vspace{-1em}
\begin{center}
\textit{Biostatistics, Bioinformatics and Epidemiology,\\ Vaccine and Infectious Diseases Division, \\ Fred Hutchinson Cancer Center}
\end{center}

\begin{abstract}
In cancer biomarker development, a key objective is to evaluate whether a new biomarker, when combined with an established one, improves early cancer detection compared to using the established biomarker alone. Incremental value is often quantified by changes at specific points on the ROC curve, such as an increase in sensitivity at a fixed specificity, which is especially relevant in early cancer detection. Our research is motivated by the Early Detection Research Network (EDRN) biorepository studies, which aim to validate multiple cancer biomarkers across laboratories using specimens obtained from a centralized biorepository, under the constraint of limited specimen availability. To address this challenge, we propose a two-stage group sequential hypothesis testing framework for assessing incremental effects, allowing early stopping for futility or efficacy to conserve valuable samples. Our asymptotic results are derived under a logistic working model and remain valid even under model misspecification, ensuring robustness and broad applicability. We further integrate a rotating group membership design to facilitate validation of multiple candidate biomarkers across laboratories. Through extensive simulations, we demonstrate valid type I error control and efficient utilization of specimens. Finally, we apply our method to data from a multicenter EDRN pancreatic cancer reference set study and show how the proposed approach identifies promising candidate biomarkers that
provide incremental performance when combined with CA19-9, while enabling efficient evaluation of a large number of such candidates.
\end{abstract}




\section{Introduction}
\label{sec1}

The development of biomarkers for cancer early detection typically follows a structured five-phase framework as described by \citet{pepe2001phases}. Phase 1 focuses on preclinical discovery studies to identify genes or proteins that are overexpressed or underexpressed in tumor tissues. In Phase 2, candidate biomarkers undergo validation with data from cancer patients and healthy controls to assess their ability to distinguish between the two groups. Phase 3 tests whether biomarkers validated in Phase 2 can effectively be used for cancer screening by evaluating their ability to detect cancer in pre-diagnostic specimens. Phase 4 assesses the performance of biomarkers in actual cancer screening scenarios, while Phase 5 examines the impact of screening on cancer-specific mortality. The methodology presented in this paper is designed for use in Phases 2 and 3, which are essential for validating promising biomarkers from initial discoveries and advancing them toward clinical utility.

Some cancers already have established biomarkers, such as CA19.9 for pancreatic cancer, PSA for prostate cancer, and alpha-fetoprotein for liver cancer, but their accuracy isn't good enough for reliable cancer detection. Assessing whether new biomarkers can add incremental value beyond existing ones is an important research problem in the early phases of biomarker validation.  Take CA 19.9 as an example. It has limited sensitivity for pancreatic cancer detection and is informative only for certain subtypes. Therefore, researchers are investigating whether incorporating additional biomarkers can improve its predictive performance and enhance overall diagnostic accuracy. Incremental performance can be evaluated using various methods, depending on the research objective. A comprehensive review of existing methods for validating incremental performance is provided in \citet{cook2018quantifying}. Likelihood-based methods assess incremental effects by leveraging assumptions of a suitable risk model that includes both established and new biomarkers. However, as noted in \citet{cook2018quantifying}, statistical significance in likelihood-based tests does not necessarily translate into clinical significance. In the context of early cancer detection, a more interpretable and clinically meaningful approach is to compare diagnostic performance directly. This is typically done by constructing two classification models—one using only the established biomarker(s), and another incorporating both established and new biomarkers—and comparing their performance metrics.  Previous literature on biomarker studies has focused on key performance metrics including sensitivity, specificity and the area under the ROC curve (AUC)  \citep{chen2013assessment, tang2008nonparametric}. However, in cancer early detection where high sensitivity is often prioritized, one of the most important metrics to look at would be the improvement in $ROC(t)$, that is, the increase in sensitivity at a pre-specified specificity of $1-t$, measured as the vertical difference between the two ROC curves. Despite its clinical relevance, incremental performance in terms of $ROC(t)$ has received limited attention in the literature. Prior work has explored $ROC(t)$ for biomarker panels \citep{tayob2016unbiased, wang2021strategies}, but not in the context of incremental testing. 

Our research is
motivated by the NCI Early Detection Research Network (EDRN) biorepository
studies, which aim to validate the performance of multiple cancer biomarkers across laboratories
using specimens from cancer cases and controls obtained from a centralized biorepository \citep{feng2013early}. One such example is the EDRN pancreatic cancer reference set study  \citep{haab2015definitive}, which collect blood samples from pancreatic cancer cases and controls for distribution to research laboratories to identify promising candidate biomarkers that compliment CA19.9. 
With multiple discovery laboratories often generating a large number of candidate markers requiring validation,
it is essential to develop a strategy that optimally utilizes the limited specimen available from the biorepository. Such an approach would enable efficient use of these valuable resources while allowing validation of more candidate markers and their incremental value. A group sequential procedure is particularly advantageous in this scenario, as it is specifically designed to maximize sample efficiency. A simple yet effective approach is the two-stage group sequential procedure for biomarker evaluation. In this design, available participants are divided into two groups, with biomarkers measured in one group first. The study proceeds to the second stage, where samples from all participants are utilized, only if certain pre-defined criteria are satisfied in the first stage. The stopping criteria in Stage 1 vary depending on the research objectives. This sequential approach helps conserve specimens and enables the validation of a greater number of biomarkers. Several recent studies have explored group sequential procedures in biomarker evaluation. For instance,  \citet{pepe2009conditional} and \citet{koopmeiners2012conditional} discussed conditional estimation of biomarker performance for a single marker, allowing for early stopping due to futility. \citet{tayob2016unbiased} and \citet{zhao2015two} extended these ideas to unbiased estimation of biomarker panel performance with a group sequential framework. More recently, \citet{wang2021strategies} developed a two-stage hypothesis testing procedure for evaluating the marginal performance of a univariate marker, incorporating a rotation scheme to optimize specimen use across multiple candidate markers. 

Building on this foundation, our work introduces a two-stage group sequential testing procedure specifically designed to assess the incremental performance of biomarkers measured from case/control samples. Our approach allows for early stopping due to futility, efficacy, or both. Moreover, it integrates a rotating group membership scheme to further optimize specimen use across multiple candidate markers. Importantly, our statistical methodology is based on minimal assumptions regarding the disease risk model, with asymptotic properties derived under a working logistic model for a panel of biomarkers, ensuring broad applicability. Previously, \citet{qin2003using, qin2010best, zhao2015two,tayob2016unbiased,park2021hybrid} discussed estimation of $ROC(t)$ for a panel assuming that a logistic regression model is correctly specified.
To our knowledge, this is the first work to combine group sequential methodology with statistical testing for incremental performance, evaluated in terms of $ROC(t)$ under a logistic working model assumption, within a case-control study framework. This extension to misspecified working models is particularly important in incremental performance assessment, where evaluating the added value of new biomarkers requires specifying both a baseline and an augmented working model in settings where multiple established markers are already included, thereby increasing the risk of model misspecification.

To demonstrate the utility of our proposed method, we apply it to data from the EDRN pancreatic cancer reference set study \citep{haab2015definitive}. This dataset includes biomarker measurements from 98 pancreatic cancer patients and 61 healthy controls, with both CA19.9 and novel biomarkers measured across all participants. Among these, a number of promising candidates have been identified. The dataset provides a realistic and challenging setting for evaluating the efficiency and effectiveness of our method in a multi-laboratory biomarker validation context and offers important guidance for optimizing sample distribution strategies in further biorepository studies. 

In summary, the main contributions of this paper are as follows. First, we develop a novel group-sequential hypothesis testing procedure to assess the incremental effects of new biomarkers added to established biomarkers, with performance measured in terms of points on the ROC curve in a case-control study setting. Second, we propose an innovative robust testing framework under a logistic working model and establish its statistical validity by deriving asymptotic properties of the test statistic while allowing for model misspecification. This contribution is broadly applicable for assessing biomarker panel performance when a convenient working risk model is adopted. 
Third, we integrate the sequential testing procedure with a rotating group framework to optimize the use of limited biorepository specimens, thereby enabling efficient assessment of the incremental performance of multiple candidate biomarkers.  

The remainder of the paper is organized as follows. Section \ref{sec:method} introduces the hypothesis testing procedure and the group rotation scheme. Section \ref{sec:sim} presents simulation studies evaluating the performance of the proposed method. Section \ref{sec:data} applies the procedure to data from a multicenter pancreatic cancer reference set study, demonstrating its practical utility. Section \ref{sec: disc} concludes with a discussion of future research directions.

\section{Methodology}
\label{sec:method}
\subsection{Objective}
\label{subsec:objective}

We consider a biorepository comprising $N_1$ cases and $N_0$ controls, with a total sample size $N=N_0+N_1$. Let $D$ denote the disease indicator where $D=1$ indicates the presence of disease and $D=0$ indicates its absence. For convenience, we also use $D_1$ and $D_0$ to refer to the diseased and non-diseased groups later in the paper. 

Suppose a panel of $m^*$ established biomarkers is already available for the disease of interest, for example, CA 19-9 for pancreatic cancer or alpha-fetoprotein for hepatocellular carcinoma. Successful validation of any new biomarker in the early phases requires demonstrating its ability to enhance diagnostic performance when combined with the established markers. Building on that, our objective is to develop statistical procedures to test whether the addition of a new biomarker to the established biomarker(s) significantly improves disease detection performance. In this work, we focus on the validation of the incremental value of a single (univariate) new marker as an initial step in biomarker validation process. However, the proposed methods are broadly applicable to evaluating the incremental contribution of multiple biomarkers jointly. 

In EDRN biorepository studies, once the specimens are ready for distribution, research laboratories interested in evaluating their biomarkers can request approval to access them. Upon approval, the corresponding specimens are distributed from the biorepository to the laboratory. Our ultimate objective is to design an optimal specimen distribution scheme that maximizes the use of available specimens, enabling the evaluation of a larger number of candidate biomarkers and facilitating the identification of more promising new biomarkers. To lay the foundation for this goal, we first describe the problem of evaluating the incremental value of a single biomarker.

Let $\theta_{r}$ denote the performance metric of interest based on established biomarkers, and $\theta_{f}$ denotes the corresponding performance metric when the new biomarker is combined with the existing ones. This is formalized as a one-sided hypothesis test: \begin{equation*} 
H_0: \theta_{f} - \theta_{r} \leq \delta_0 \text{ vs } H_1: \theta_{f} -\theta_{r}> \delta_0 \end{equation*} where $\delta_0>0$ is some pre-specified threshold value, and the test is conducted at a significance level $\alpha$. 

In this paper, we focus on the performance metric $\theta= ROC(t)$,  defined as the true positive fraction (sensitivity) at a fixed false positive fraction $t$ (that is, a specificity of $1-t$). Specifically, $\theta_f= ROC_f(t)$ and $\theta_r = ROC_r(t)$ represent the sensitivity at false positive rate $t$ for the models using the combined panel and only the established biomarkers, respectively.

The conventional strategy involves testing the hypothesis using samples from all participants collectively. While straightforward, this approach may not effectively utilize scarce resources such as blood or cystic fluid. To address this,  we propose a two-stage group-sequential testing strategy for validating the incremental performance of a new biomarker.
This approach enhances resource efficiency by incorporating an early stopping mechanism when preliminary results are conclusive, thereby streamlining the testing process and reducing unnecessary specimen consumption.

Let $Z_k$ denote the test statistic at the $k^{th}$ stage, where $k=1,2$. The sequential testing proceeds as follows: \begin{align*}
   & \text{If } Z_1 \geq b_1, \text{ Reject } H_0 \text{ (Early efficacy)}\\ 
    & \text{If } Z_1 \leq a_1, \text{ Accept } H_0 \text{ (Early futility)}\\ 
     & \text{If } a_1 < Z_1 < b_1, \text{ Continue to the next stage } \\ 
      & \text{If } Z_2 \geq b_2, \text{ Reject } H_0 \\ 
       & \text{If } Z_2 < a_2, \text{ Fail to reject } H_0 
\end{align*} Here $a_1 <b_1$, and $a_2=b_2$ ensures termination by the second stage. The selection of the boundaries $a_k$, $b_k$ is discussed in detail in the following subsection.

\subsection{Determination of critical values for one-sided symmetric two-stage group sequential test}
\label{subsec:cutoff}
Let $Z_1$ and $Z_2$ denote the test statistics computed at the first and second stages of the analysis, with corresponding information levels $I_1$, $I_2$. For a two-stage equally spaced design, it is common to assume $I_1= \frac{1}{2}I_2$ \citep{jennison1999group}. To test the hypothesis specified in Section \ref{subsec:objective} at a significance level $\alpha$, we must have \begin{equation*}
    P_{H_0}(\text{Reject } H_0) = \alpha.
\end{equation*} This can be decomposed as: 
\begin{equation*}
      P_{H_0}(Z_1 \geq b_1)+P_{H_0}(a_1 < Z_1 < b_1, Z_2 \geq b_2) = \alpha 
\end{equation*} where $a_1 < b_1$, $a_2=b_2$ to ensure termination by the second stage.
To determine the cutoffs $a_1$, $b_1$ and $b_2$, we can implement the $\alpha$ spending approach proposed by \citet{gordon1983discrete}. Let $\alpha_1$ and $\alpha_2$ be two probabilities with $\alpha_1+\alpha_2=\alpha$. Then, we have the following equations: 
\begin{equation}
\label{eq1}
    P(Z_1 \geq b_1) = \alpha_1,
\end{equation} and \begin{equation}
\label{eq2}
    P(a_1 < Z_1 < b_1, Z_2 \geq b_2) = \alpha_2.
\end{equation}
As \citet{gordon1983discrete} suggests, $\alpha_1$ and $\alpha_2$ can be determined on the basis of an error spending function $f(t)$, a non-decreasing function satisfying $f(0)=0$, and $f(1)=\alpha$. Then, we can choose \begin{equation*}
    \alpha_1 = f(I_1/I_2) \text{ and } \alpha_2 = \alpha- \alpha_1.
\end{equation*} 
Suppose $(Z_1, Z_2)$ follows the canonical joint distribution, that is, they satisfy the following set of conditions. \begin{enumerate}
    \item $(Z_1, Z_2)$ is bivariate normal.
    \item $E(Z_k)= \delta \sqrt{I_k}$ for $k=1,2$. for some $\delta$.
    \item $\text{Cov}(Z_1, Z_2)=\sqrt{I_1/I_2}$.
\end{enumerate}
Under this assumption of a canonical joint distribution, the following functions provide approximations to the classical O'Brien-Fleming boundary \citep{o1979multiple} and Pocock boundary \citep{pocock1977group} respectively: \begin{equation*}
    f(t) = 2 \Phi\left (-\frac{z_{\alpha/2}}{\sqrt{t}}\right),
\end{equation*}
and
\begin{equation*}
    f(t) = \alpha \log(1+(e-1)t).
\end{equation*}
Note that Equations (\ref{eq1}) and (\ref{eq2}) are not sufficient to determine all the cutoffs. Hence, we adopt the symmetric group-sequential framework \citep{emerson1989symmetric, wassmer2016group,tamhane2021group}. For this, we equate the Type 2 error under the alternative $\delta=\delta_1$ ($>\delta_0$, denoted by $\beta$) to the corresponding Type 1 error $\alpha$. That is, we consider $\beta=\alpha$. This implies, \begin{equation*}
    P_{H_1}(\text{Accept } H_0) = \alpha,
\end{equation*} 
with the following conditions: \begin{equation*}
\label{eq3}
    P_{H_1}(Z_1 \leq a_1) = \alpha_1,
\end{equation*} and \begin{equation*}
\label{eq4}
    P_{H_1}(a_1 < Z_1 <  b_1, Z_2 \leq a_2) = \alpha_2
\end{equation*} with $\alpha_1+\alpha_2=\alpha$. Due to the symmetry of the design,it follows that, \begin{equation}
\label{eq5}
    a_k = \delta_1 \sqrt{I_k}- b_k
\end{equation} for $k=1,2$. Since the procedure terminates at Stage 2, that is $a_2=b_2$, we obtain:
\begin{equation*}
    b_2 = \frac{\delta_1}{2} \sqrt{I_2}
\end{equation*} which implies \begin{equation*}
    \delta_1 = \frac{2 b_2}{\sqrt{I_2}}
\end{equation*} Substituting the value of $\delta_1$ in Equation (\ref{eq5}) for $k=1$, we get, \begin{equation*}
    a_1 = 2 b_2 \sqrt{\frac{I_1}{I_2}} -b_1.
\end{equation*}
For the commonly used design where $I_1/I_2 =1/2$, this simplifies to $a_1= \sqrt{2}b_2 -b_1$.

This framework applies to designs allowing early stopping for both futility and efficacy. If we want the procedure to stop for efficacy only, we set $a_1= -\infty$. Similarly, if we want the procedure to stop for futility only, we set $b_1= \infty$.

\subsection{Definition and Estimation of ROC(t)}
\label{subsec:estimation}

Before introducing the testing procedure for our hypothesis testing procedure, we first describe how we define and estimate $ROC(t)$ for a panel of biomarkers with an appropriate choice of risk model. Throughout this paper, we frequently consider two sets of biomarker panels, one including and the other excluding the new marker. Without loss of generality, we focus on evaluating the incremental value of adding one new biomarker, as is commonly done in biomarker validation. However, the methods we develop apply more broadly to settings where multiple new markers are assessed jointly for their incremental value. 

We refer to the panel containing all biomarkers as the `full panel',  and the panel without the new biomarker as the `restricted panel'. Accordingly, we refer to the associated risk models as the `full model' and `restricted model', respectively. Since we employ a two-stage group sequential procedure, we use $k=1,2$ to denote the stages. Further, we define the biomarker vectors for the full and restricted panels as ${X}^f = (X_{1}, \ldots, X_{(m^*+1)})$ and ${X}^r = (X_{1}, \ldots, X_{m^*})$, representing the $m^*+1$ and $m^*$ biomarker values, respectively.

  As noted in \citet{mcintosh2002combining} and \citet{zhao2015two}, the optimal approach for combining a set of biomarkers to maximize the ROC curve is to use the true disease risk conditional on the biomarker set, $p()$, or its monotone transformation. In practice, a working model for disease risk is typically assumed in order to derive the marker combination.  Suppose $ROC_f(t)$ and $ROC_r(t)$ represent the ROC curves based on the full and restricted working models, respectively. We denote their corresponding estimators as $\widehat{ROC}_f(t)$ and $\widehat{ROC}_r(t)$, obtained using an appropriate estimation method.

\subsubsection{Logistic Working model}
 We adopt logistic regression as a working model for the disease risk given each biomarker panel. That is, we adopt a working model given the full panel as
$1/\left(1+e^{-(\alpha_f+X^{f'}\beta_f)}\right)$
for some intercept parameter $\alpha_f$ and some slope parameter $\beta_f$. Similarly, we adopt working model given the restricted panel as
$1/\left(1+e^{-(\alpha_r+X^{r'}\beta_r)}\right)$
for some intercept parameter $\alpha_r$ and some slope parameter $\beta_r$.

  Let $\alpha_f^*$, $\alpha_r^*$, $\beta_f^*$ and $\beta_r^*$ denote the pseudo-true intercept and slope parameters, defined as unique solutions to the following equations \begin{align*}
    E_{cc} \left( X^f \left(D- \frac{e^{\alpha_f+{X}^{f'}\beta_f}}{1+e^{\alpha_f+X^{f'}\beta_f}} \right)\right)= 0 \\
      E_{cc} \left( X^r \left(D- \frac{e^{\alpha_r+{X}^{r'}\beta_r}}{1+e^{\alpha_r+X^{r'}\beta_r}} \right)\right)= 0 \\
\end{align*} where $E_{cc}(.)$ denotes expectation under the mixture distribution of case and control populations, where the case-to-control ratio is taken to be the same as in the sample. Equivalently, these equations can be written as \begin{align*}
   \pi_{case} E_{case} \left( X^f \left(1- \frac{e^{\alpha_f+{X}^{f'}\beta_f}}{1+e^{\alpha_f+X^{f'}\beta_f}} \right)\right)- (1-\pi_{case}) E_{con} \left( X^f \left(\frac{e^{\alpha_f+{X}^{f'}\beta_f}}{1+e^{\alpha_f+X^{f'}\beta_f}} \right)\right)= 0  
\end{align*} and \begin{align*}
    \pi_{case} E_{case} \left( X^r \left(1- \frac{e^{\alpha_r+{X}^{r'}\beta_r}}{1+e^{\alpha_r+X^{r'}\beta_r}} \right)\right)- (1-\pi_{case}) E_{con} \left( X^r \left(\frac{e^{\alpha_r+{X}^{r'}\beta_r}}{1+e^{\alpha_r+X^{r'}\beta_r}} \right)\right)= 0  
\end{align*} where $E_{case}(.)$ and $E_{con}(.)$ denote expectations taken with respect to the conditional distributions of $X|D=1$ and $X|D=0$ respectively, and $\pi_{case}$ denotes the proportion of cases in the sample. 

When the logistic regression model correctly specifies the disease risk in the underlying general population from which the case–control sample is obtained, the maximum likelihood estimators based on the case–control sample are consistent for the model parameters (up to an intercept shift) \citep{prentice1979logistic}, where these estimators of slope parameters are invariant to the case-control ratio. However, when the working logistic model is misspecified, as is often the case in practice, the pseudo-true slope parameters depend on the case-control ratio in the mixture distribution. Consequently, the logistic regression slope estimators converge to different pseudo-true values when the case-to-control ratio changes \citep{xie1989logit}. In two-phase sampling studies,weighted likelihood methods that weight cases and control by the inverse of their sampling probabilities (\citet{manski1977estimation,xie1989logit}) can remove this dependence by reweighting the data to mimic the underlying population from which cases and controls are drawn. In this paper, we adopt standard unweighted logistic regression when only a representative case–control sample is available, as in our pancreatic cancer biorepository study. We note, however, that methods incorporating sampling weights in two-phase sampling settings represent a valuable direction for future methodological extension.

Let $\hat{\beta}_{kf}$ and $\hat{\beta}_{kr}$ denote the maximum likelihood estimators of the slope parameters for the full and restricted models, respectively, obtained by fitting the logistic regression working model to the case-control data at step $k$ of the testing procedure. These estimators solve the sample score equations, and under standard regularity conditions, they converge to the pseudo-true parameters $\beta_f^*$ and $\beta_r^*$ for a fixed case-control ratio.

\subsubsection{Population ROC curve}
Let $D_1$ and $D_0$ denote cases and controls, respectively, used interchangeably with $D=1$ and $D=0$. For given slope vectors $\beta_f$ and $\beta_r$, suppose the biomarker combination scores $X_{}^{f'}\beta_f|D_1$ and $X_{}^{f'}\beta_f|D_0$ follow continuous distributions with distribution functions denoted by $F_{D_1,f}(\beta_f,.)$ and $F_{D_0,f}(\beta_f,.)$. Similarly, $X_{}^{r'}\beta_r|D_1$ and $X_{}^{r'}\beta_r|D_0$ have distributions denoted by $F_{D_1,r}(\beta_r,.)$ and $F_{D_0,r}(\beta_r,.)$. 

Then, the population ROC curves induced by the combination scores are defined as 

\begin{equation*}
    ROC_{f}(t, \beta_f) = 1- F_{D_1,f}(\beta_f,F_{D_0,f}^{-1}(\beta_f,1-t))
\end{equation*} and \begin{equation*}
     ROC_{r}(t, \beta_r) = 1- F_{D_1,r}(\beta_r,F_{D_0,r}^{-1}(\beta_r,1-t))
\end{equation*} 

With a little abuse of notation, we denote the ROC curves at pseudo true parameters defined above as \begin{equation}
      ROC_{f}(t) \equiv   ROC_{f}(t, \beta_f^*)
\end{equation}

\begin{equation}
      ROC_{r}(t) \equiv   ROC_{r}(t, \beta_r^*).
\end{equation}

\subsubsection{Estimation of $ROC_f(t)$ and $ROC_r(t)$}

Let $X_{ij}$ denote the value of the $j^{th}$ biomarker measured for the $i^{th}$ individual, where $j=1, \ldots, (m^*+1)$ and $i=1, \dots, N$ with $N=N_1+N_0$. Further, let $\lambda_{D_1}$ and $\lambda_{D_0}$ denote the fraction of cases and controls chosen at stage $k$. For equal fractions of samples from cases and controls at stage 1, we set $\lambda_{D_1}=\lambda_{D_0}=\lambda$ for $k=1$ and $\lambda_{D_1}=\lambda_{D_0}=1$ for $k=2$.

The empirical estimators of $ROC(t)$ at step $k$ are then given by: \begin{equation}
  \text{Full: }  \widehat{ROC}_{kf}(t) = \widehat{S}_{D_1, f}(\hat{\beta}_{kf},\widehat{S}^{-1}_{D_0, f}(t)),
\end{equation} \begin{equation}
    \text{Restricted: }  \widehat{ROC}_{kr}(t) = \widehat{S}_{D_1, r}(\hat{\beta}_{kr},\widehat{S}^{-1}_{D_0, r}(t)), 
\end{equation} where \begin{equation*}    
\widehat{S}_{D_1,f}(\hat{\beta}_{kf},t)= \frac{1}{[\lambda_{D_1}N_1]} \sum_{i=1}^{[\lambda_{D_1}N_1]} I (X_i^{f'}\hat{\beta}_{kf}>t, D=1) \ \text{, for } \frac{1}{N_1} \leq \lambda_{D_1} \leq 1,\end{equation*}, \begin{equation*}    
\widehat{S}_{D_1, r}(\hat{\beta}_{kr},t)= \frac{1}{[\lambda_{D_1}N_1]} \sum_{i=1}^{[\lambda_{D_1}N_1]} I (X_i^{r'}\hat{\beta}_{kr}>t, D=1) \ \text{, for } \frac{1}{N_1} \leq \lambda_{D_1} \leq 1,\end{equation*} 
\begin{equation*}    
\widehat{S}^{-1}_{D_0,f}(\hat{\beta}_{kf},1-t)= \begin{cases} & X_{[1]}^{f'}\hat{\beta}_{kf}, \ \text{ for } t=0, D=0 \\ & X_{[s]}^{f'}\hat{\beta}_{kf}, \ \text{ for } \frac{s-1}{[\lambda_{D_0}N_0]} < t \leq \frac{s}{[\lambda_{D_0}N_0]}, D=0, \end{cases}\end{equation*}
\begin{equation*}    
\widehat{S}^{-1}_{D_0, r}(\hat{\beta}_{kr},1-t)= \begin{cases} & X_{[1]}^{r'}\hat{\beta}_{kr}, \ \text{ for } t=0, D=0 \\ & X_{[s]}^{r'}\hat{\beta}_{kr}, \ \text{ for } \frac{s-1}{[\lambda_{D_0}N_0]} < t \leq \frac{s}{[\lambda_{D_0}N_0]}, D=0, \end{cases}\end{equation*}
where $(X_{[1]}^{f'}\hat{\beta}_{kf}, \ldots, X_{[\lambda_{D_0}N_0]}^{f'}\hat{\beta}_{kf})$, $(X_{[1]}^{r'}\hat{\beta}_{kr}, \ldots, X_{[\lambda_{D_0}N_0]}^{r'}\hat{\beta}_{kr})$ denote the ordered biomarker combination scores among controls for the full and restricted models respectively.


To develop a two-stage sequential testing procedure for assessing the incremental value of a new biomarker in terms of $ROC(t)$, we begin by establishing a suitable statistic. A natural choice is the difference in estimated sensitivities at a fixed false positive rate $t$, computed from the full and restricted models. Specifically, at stages $k \in \{1,2\}$, we consider the quantity  $$\widehat{ROC}_{kf}(t)-\widehat{ROC}_{kr}(t).$$ For simplicity, we focus on Stage 1 and omit the subscript $k$, with analogous results holding for Stage 2. To construct a valid hypothesis test, we require the asymptotic distribution of the above incremental value estimate, which forms the basis for our group sequential test statistic. 

We now state an assumption that ensures that the quantile functional is differentiable and the kernel density estimator used for plug-in variance estimation later in our asymptotic results remains consistent.

\begin{assumption}[Smoothness]
\label{ass2}
The distribution of the control biomarker combination scores $X_i^{f'}\beta_f^*|D_0$ and $X_i^{r'}\beta_r^*|D_0$ are continuous with strictly positive densities at the $(1-t)^{th}$ quantile. 
    
\end{assumption} 

\noindent \textbf{Asymptotic Distribution of ROC Estimator}

\noindent Before stating the theorem on the asymptotic distribution of the incremental $ROC$, we state another theorem which forms a foundational result on the asymptotic distribution of $ROC(t)$ under the logistic regression working model and a fixed case-control ratio.

\begin{theorem}
\label{theorem:th1}
Let $\widehat{ROC}_f(t)$ denote the estimator of $ROC_f(t)$ based on the procedure described in Section \ref{subsec:estimation}. Then, under Assumption \ref{ass2}, and for sufficiently large sample sizes with a fixed case-control ratio, $\widehat{ROC}_f(t)$ is asymptotically normally distributed with mean $ROC_f(t)$ and variance given by $\Sigma_f$.
\end{theorem}

Here, $\Sigma_f$ denotes the variance of the estimator of $ROC_f(t)$ with the explicit expression for $\Sigma_f$, along with the proof of this theorem, provided in Appendix A of the Supplementary Material. While we retain the subscript $f$ for notational consistency, it is important to note that the result holds generally for $ROC(t)$ when estimated under a logistic regression working model for a panel of biomarkers.

\noindent \textbf{Asymptotic Distribution of Incremental ROC Difference}

\noindent We now state our main result that establishes the asymptotic normality of $\widehat{ROC}_{f}(t)-  \widehat{ROC}_{r}(t)$ under assumption 1.

\begin{theorem} 
\label{theorem:th2}
Let $\widehat{ROC}_f(t)$ and $\widehat{ROC}_r(t)$ denote the estimators of $ROC_f(t)$ and $ROC_r(t)$. Then, under Assumption \ref{ass2}, for sufficiently large sample sizes and for a fixed case-control ratio, the difference $$\widehat{ROC}_f(t)-\widehat{ROC}_r(t)$$ is asymptotically normal with mean $ROC_f(t)-ROC_r(t)$ and variance given by $ \Sigma_f +\Sigma_r -2 \Sigma_{fr}$.

\end{theorem}

Here, $\Sigma_f$, $\Sigma_r$, and $\Sigma_{fr}$ denote the variances of the estimators of $ROC_f(t)$, $ROC_r(t)$, and their covariance respectively. The proof of Theorem \ref{theorem:th2} can be found in Appendix B of the Supplementary Material. The explicit expressions for the quantities $\Sigma_{f}$, $\Sigma_{r}$, and $\Sigma_{fr}$ can be found in Appendices A and B. This theorem establishes the asymptotic normality of the difference in estimated sensitivities at a specificity of $1-t$. This result also provides the theoretical foundation for constructing a standardized test statistic for each stage of the sequential procedure, which will be discussed in the next section. While for the sake of simplicity, we focus on the incremental effect of adding a single new biomarker to the panel, it is worth noting that the asymptotic result holds in a more general setting and is applicable even when multiple biomarkers are incorporated into the existing panel.

\subsection{Sequential hypothesis testing framework for evaluating the incremental value}

\label{subseq:seq_test}

We now return to the primary objective: testing whether the addition of a new biomarker significantly improves sensitivity at a fixed specificity level $1-t$. Formally, we test the following one-sided hypothesis: \begin{equation*}
 H_0: ROC_f(t)-ROC_r(t) \leq \delta_0 \ \ vs \ \  H_1: ROC_f(t)-ROC_r(t) > \delta_0
\end{equation*} where $\delta_0>0$ is a pre-specified threshold representing the minimum clinically meaningful improvement, and the test is conducted at significance level $\alpha$. Let $\Sigma=  \Sigma_f +\Sigma_r -2 \Sigma_{fr}$ denote the asymptotic variance of the difference in estimated ROC values, as established in Theorem \ref{theorem:th2}. Then, we use the following as our test statistic in Stage 1: \begin{equation*}
   Z_1= \frac{[\widehat{ROC}_f(t)-\widehat{ROC}_r(t)]-\delta_0}{\sqrt{\widehat{\Sigma}} }
\end{equation*} where $\widehat{\Sigma}$ is a plug-in estimator of $\Sigma$.  We note from the expression of $\Sigma$ in Appendix B that $\Sigma$ comprises distribution functions and density functions. In our plug-in estimator, we replace the distribution functions by the corresponding empirical distribution functions, and the densities by the corresponding kernel density estimators. 

Similarly, let $Z_2$ denote the test statistic obtained in Stage 2, constructed analogously using data from both stages. 

To implement the one-sided symmetric test design as discussed in \cite{emerson1989symmetric, wassmer2016group, tamhane2021group}, we must have, \begin{equation*}
    P_{H_0}(\text{Reject } H_0) \leq \alpha
\end{equation*} and \begin{equation*}
    P_{H_1}(\text{Accept } H_0) \leq \alpha
\end{equation*}
Assuming that $(Z_1, Z_2)$ follows the canonical joint distribution (as discussed in Section \ref{subsec:cutoff}), the critical boundaries $a_1, b_1, a_2=b_2$ are determined using the $\alpha$-spending approach and the symmetry conditions.

\subsection{Rotation of group membership to optimize biorepository specimen use}
\label{subsec:rotation}

In a biorepository study with limited specimen availability to evaluate multiple biomarkers from various laboratories, our goal is to assess as many candidate biomarkers as possible for their incremental value beyond established markers. A notable example is the EDRN pancreatic cancer reference set study, in which blood samples were collected to validate pancreatic cancer biomarkers \citep{haab2015definitive}. In this study, statisticians from the Data Monitoring and Coordinating Center distribute blood specimens to participating laboratories to assess each biomarker's incremental value over CA 19-9. 

When using a two-stage sequential procedure, a fixed allocation of participants to Stage 1 and Stage 2 across all biomarkers can lead to rapid depletion of specimens from individuals assigned to Stage 1. As a result, only specimens from the remaining participants will be available for biomarkers analyzed later in the sequence. To address this, we implement a group rotation scheme, originally proposed by \citet{wang2021strategies} for evaluating the marginal performance of candidate markers. Here, we adapt the approach to assess incremental value, making efficient use of specimens across all participants and biomarkers. The algorithm is outlined below. 

\begin{algorithm}[H]
\caption{Group rotation for validating incremental effect of biomarkers}\label{alg:cap}
\begin{itemize}
   \item In the biorepository, suppose we have $V+m^*$ units of specimens from $N_1$ cases and $N_0$ controls
    \item Without loss of generality, assume one unit of specimen is used for measuring one biomarker. Initially, we use the first $m^*$ units to measure the $m^*$ established biomarkers.
    \item We partition the participants into $\kappa=1/\lambda$ groups, stratified by case/control status. 
    \item When a new biomarker comes in for validation of incremental value
    \begin{itemize}
    \item \textbf{While} at least one unit of specimen remains in individuals across all groups, we
    \begin{itemize}
       \item randomly select one group from among the groups with the highest number of units left to serve as the stage 1 sample.
       \item combine the remaining groups to form the Stage 2 sample.
       \item measure the biomarker for the stage 1 sample and conduct the hypothesis test for incremental value over the established markers. If the test proceeds to stage 2, measure biomarkers for all samples and conduct the test.
       \end{itemize}
   \item \textbf{If} some groups have some units left, we can
   \begin{itemize}
       \item  conduct fixed sample test using those.
   \end{itemize}
    \end{itemize}
    \end{itemize}
\end{algorithm}

This rotation strategy ensures that specimens are distributed evenly across biomarkers and participants, maximizing the number of biomarkers for which the incremental value can be evaluated. It also allows for adaptive decision-making based on specimen availability, making it particularly suitable for large-scale biomarker validation.


Given a total of $V+m^*$ available specimen units in the biorepository, we derive analytical expressions for key performance metrics of the proposed rotating group sequential design. These include, 
(i) the expected number of biomarkers measurable for assessing incremental value (Result \ref{res1});
(ii) the expected number of biomarkers identified as having significant incremental value (Result \ref{res2}); and
(iii) the expected number of truly useful biomarkers that can be successfully validated (Result \ref{res3}).

\begin{result}[\textbf{Expected Number of Biomarkers Evaluated}]
\label{res1} Let $n^*$ denote the number of biomarkers whose incremental performance over the established biomarker is to be evaluated. The expected number of biomarkers that can be evaluated is given by the following formula: \begin{align*}
        E(n^*) &= V+ \sum_{i=0}^V (\kappa-1)i {V + (\kappa-1)i \choose \kappa i} p^{\kappa i}(1-p)^{V-i} \\
        &+ \sum_{i=0}^{V-1} \sum_{j=0}^{\kappa-2} ((\kappa-1)i+j) {V + (\kappa-1)i + j \choose \kappa i+j+1} p^{\kappa i+j+1}(1-p)^{V-i}
    \end{align*} where $p=P(Z_1\geq b_1)+P(Z_1 \leq a_1)$.
\end{result}

\begin{result}[\textbf{Expected Number of Biomarkers Identified as Significant}]
\label{res2}
    Let $n_u^*$ denote the number of biomarkers for which the null hypothesis rejected at any stage. Its expected value is given by the following formula \begin{equation*}
        E(n_u^*)=E(n^*)p_r
    \end{equation*} where $p_r=P(Z_1\geq b_1)+P(a_1 \leq Z_1 \leq b_1, Z_2 \geq b_2)$ is the probability of rejection.
\end{result}

\begin{result}[\textbf{Expected Number of Truly Useful Biomarkers Successfully Validated}]
\label{res3}
Suppose the true performance of the biomarkers to be validated in the reference set follows a mixture distribution, i.e.,
$\theta_f-\theta_r = \theta^*\le \delta_0$ with probability $\gamma$ (null marker) and $\theta_f-\theta_r=\delta_A>\delta_0$ with probability $1-\gamma$ (truly useful marker), where $\gamma \in [0, 1]$. 
    Let $n_u^{t*}$ denote the number of truly useful biomarkers that are successfully validated. Its expected value is given by the following formula, \begin{equation*}
        E(n_u^{t*})= E(n^*)p_r^* (1-\gamma)
    \end{equation*} where $p_r^*= P_{H_1}(Z_1\geq b_1)+P_{H_1}(a_1 \leq Z_1 \leq b_1, Z_2 \geq b_2)$.
\end{result}
These analytical formulas can provide preliminary insights for assessing the performance of the proposed designs, taking into account various parameters to optimize the use of biorepository specimens.

\section{Simulation study} 
\label{sec:sim}
\subsection{Hypothesis Testing} We conduct simulation studies to evaluate the performance of the proposed hypothesis testing procedure for assessing incremental biomarker value under two scenarios: (i) when the logistic working model is correctly specified and (ii) when it is misspecified. Below, we describe the simulation settings and results.

\subsubsection{Hypothesis testing framework}

Our goal is to test whether adding a new biomarker improves sensitivity at a fixed specificity level. Specifically, we consider:

\begin{equation*}
  H_0: ROC_f(0.1)-ROC_r(0.1) \leq \delta_0 \ \ \text{vs} \ \ H_1: ROC_f(0.1)-ROC_r(0.1) > \delta_0
\end{equation*} where $\delta_0$ is a clinically meaningful threshold. We set $\delta_0 = 0.141$ for the correctly specified model and $\delta_0 = 0.165$ for the misspecified model, corresponding to a null scenario to be described in the following subsections.

\subsubsection{Risk model correctly specified}
Let $X_1$ and $X_2$ denote the established biomarker and new candidate biomarker, respectively. We begin by generating case-control data for the `full model', which include both biomarkers $X_1$ and $X_2$, such that \begin{equation*}
   X^{f}|D=0 \sim MVN(0, \Sigma_{{f,sim}})
\end{equation*} and \begin{equation*}
    X^{f}|D=1 \sim MVN( \mu_{f}, \Sigma_{f,sim})
\end{equation*} where $X^{f}=(X_1,X_2)$, $\mu_{f}=(\mu_1, \mu
_2)$, and $\Sigma_{f,sim}$ is a covariance matrix in which each marker has a variance of 1, and the covariance between the two biomarkers is 0.2. For this model, the optimal risk score $P(D=1|X^{f})$ is a monotone increasing function of $X^{f'}\beta_f$, where $\beta_f= \Sigma_{f,sim}^{-1}\mu_f$. In this scenario, the ROC curve, $ROC_f(t)$, has a closed-form expression given by \begin{equation*}
    ROC_{f}(t) = \Phi\left(\sqrt{\mu_{f}'\Sigma_{f,sim}^{-1}\mu_{f}}+\Phi^{-1}(t)\right)
\end{equation*} The `restricted model' is obtained by removing $X_2$, leaving only $X_1$. Thus, we have, 
   \begin{equation*}
   X^{r}|D=0 \sim MVN(0, \Sigma_{r,sim})
\end{equation*} and \begin{equation*}
    X^{r}|D=1 \sim MVN( \mu_{r}, \Sigma_{r, sim})
\end{equation*} where $X^{r}=X_1$, $\mu_{r}=\mu_1$, and $\Sigma_{r, sim}=1$. The corresponding ROC curve, $ROC_{r}(t)$, follows a similar form as above. 

Suppose we are interested in $t=0.1$, where the incremental value of the full model over the restricted model is measured by the difference in sensitivities at a specificity level of 0.9. In these simulation studies, we fix $\mu_r=1$ so that the established biomarker has $ROC_r(0.1)=0.389$. Choosing $\mu_f= (1, 1.1)$ corresponds to $ROC_f(0.1)-ROC_r(0.1)=0.141$ matching $\delta_0$. We also consider alternative scenarios with $\mu_f= (1,1.5)$ and $(0.8,2)$ corresponding to larger incremental values $\delta_1= 0.259$ and $0.461$ respectively.


\subsubsection{Risk model misspecified}
We now consider a model similar to above, with the exception that we now allow the covariance matrices to vary between the cases and controls. More specifically, we assume \begin{equation*}
   X^{f}|D=0 \sim MVN(0, \Sigma_{D_0f,sim})
\end{equation*} and \begin{equation*}
    X^{f}|D=1 \sim MVN( \mu_{f}, \Sigma_{D_1f,sim})
\end{equation*} where $X^{f}=(X_1,X_2)$, $\mu_{f}=(\mu_1, \mu
_2)$, and $\Sigma_{D_1f,sim}$ and $\Sigma_{D_0f,sim}$ are covariance matrices for the cases and controls in which each marker has a variance of 1, and the covariances between the two biomarkers are 0.2 and 0.1, respectively. In this case, the linear logistic risk model is misspecified and we do not get a closed-form expression for $ROC_f(t)$. The hypothesis and simulation design mirror the previous scenario, except $\delta_0 = 0.165$ corresponding to $\mu_f=(1,1.1)$. Under alternative scenarios with $\mu_f= (1,1.5)$ and $(0.8,2)$ we have $\delta_1= 0.278$ and $0.467$ respectively.


\subsubsection{Results}
We now present the estimated probabilities of rejection at Stages 1 and 2, as well as the probability of continuation to stage 2 under both the null and alternate scenarios. We consider two sample size scenarios: one with 200 cases and 200 controls, and another with 200 cases and 400 controls. In each scenario, we used 5000 Monte Carlo replicates with $\lambda=1/2$. Tables \ref{tab:prob_0.1_uneq} and \ref{tab:prob_0.1_fut_uneq} present the results based on the misspecified model. Results for the correctly specified model, which show similar patterns, appear in Tables \ref{tab:prob_0.1} and \ref{tab:prob_0.1_fut} in Appendix C.1. of the Supplementary material. In Tables \ref{tab:prob_0.1_uneq} and \ref{tab:prob_0.1_fut_uneq}, we make use of the asymptotic normal distribution of Theorem \ref{theorem:th2} to determine the cutoffs in both stages, applying the O' Brien and Fleming and Pocock boundaries. Table \ref{tab:prob_0.1_uneq}  allows stopping in Stage 1 for both futility and efficacy. Table \ref{tab:prob_0.1_fut_uneq} follows a similar approach but allows stopping only for futility. These tables confirm the validity of our proposed testing procedures. The Type-I error rates are close to nominal levels for both O' Brien Fleming and Pocock boundaries, regardless of the stopping criteria used. Furthermore, as the incremental value deviates further from the null, the probability of rejection approaches 1, which is expected. Note that, using O' Brien Fleming boundaries leads to a higher probability of continuation to Stage 2 compared to the Pocock boundaries, as the former have wider thresholds. However, this also implies that using O' Brien and Fleming boundaries would lead to a greater sample usage for biomarker validation compared to the Pocock boundaries. Additionally, stopping only for futility (as opposed to both futility and efficacy) increases the total number of samples required for biomarker validation. As a result, fewer biomarkers are likely to be validated for their incremental performance. If $\delta_1$ is much larger than $\delta_0$ and we are stopping only for futility, the probability of continuing to Stage 2 approaches 1, implying that nearly all available samples will be used up, as can be seen from Table \ref{tab:prob_0.1_fut_uneq}. The tables in the Appendix, based on the correctly specified model, show similar patterns. There, we observe that the type 1 error is close to the nominal levels, and as the incremental value deviates from the null, the probability of rejection approaches 1, as we would expect.

\begin{table}[tbp]
  \caption{Probabilities of rejection in Stage 1, Stage 2, combined probabilities of rejection, and the probabilities of continuing to Stage 2 for the misspecified model and for two boundary choices: O'Brien Fleming \& Pocock, with 200 cases, 200 controls; and 200 cases, 400 controls for $t=0.1, \lambda= 1/2 \ \& \   \delta_0=0.165$.  Here, we stop for both futility and efficacy.}
    \centering
    \resizebox{\textwidth}{!}{
        \begin{tabular}{cccccccc}
            \hline
            Boundary & Case & $\mu_{f}$ & $\delta$ & $S1: P(Reject|H)$ & $S1: P(Continue|H)$ & $S2: P(Reject|H)$ & $P(Reject|H)$ \\ \midrule
          \multicolumn{8}{c}{200 cases and 200 controls} \\ \hline
            \multirow{3}{*}{O'Brien Fleming} 
            & Null & $(1,1.1)$ & 0.165 & 0.015 (0.002) & 0.483 (0.007) & 0.042 (0.003) & 0.057 (0.003) \\ 
            & \multirow{2}{*}{Alternate} 
            & (1,1.5) & 0.278 & 0.146 (0.005) & 0.753 (0.006) & 0.396 (0.007) & 0.542 (0.007) \\ 
            &  & (0.8, 2) & 0.467 & 0.825 (0.005) & 0.175 (0.005) & 0.173 (0.005) & 0.998 (0.001) \\ \hline
            
            \multirow{3}{*}{Pocock} 
            & Null & $(1,1.1)$ & 0.165 & 0.036 (0.003) & 0.179 (0.005) & 0.016 (0.002) & 0.052 (0.003) \\ 
            & \multirow{2}{*}{Alternate} 
            & (1,1.5) & 0.278 & 0.299 (0.006) & 0.388 (0.007) & 0.180 (0.005) & 0.479 (0.007) \\ 
            &  & (0.8,2) & 0.467 & 0.931 (0.004) & 0.063 (0.003) & 0.060 (0.003) & 0.991 (0.001) \\ \hline
            \multicolumn{8}{c}{200 cases and 400 controls} \\ \hline  
            \multirow{3}{*}{O'Brien Fleming} 
            & Null & $(1,1.1)$ & 0.165 & 0.010 (0.001) & 0.485 (0.007) & 0.038 (0.003) & 0.048 (0.003) \\ 
            & \multirow{2}{*}{Alternate} 
            & (1,1.5) & 0.278 &0.202 (0.006) & 0.743 (0.006) & 0.492 (0.007) & 0.694 (0.007) \\ 
            &  & (0.8, 2) & 0.467 & 0.932 (0.004) & 0.053 (0.003) & 0.053 (0.003) & 1.000 (0.000) \\ \hline
            
            \multirow{3}{*}{Pocock} 
            & Null & $(1,1.1)$ & 0.165 & 0.031 (0.002) & 0.175 (0.005) & 0.015 (0.002) & 0.047 (0.003) \\ 
            & \multirow{2}{*}{Alternate} 
            & (1,1.5) & 0.278 & 0.368 (0.007) & 0.405 (0.007) & 0.233 (0.006) & 0.601 (0.007) \\ 
            &  & (0.8, 2) & 0.467 & 0.983 (0.002) & 0.016 (0.002) & 0.016 (0.002) & 0.999 (0.000) \\ \hline
        \end{tabular}
    }
    \label{tab:prob_0.1_uneq}
\end{table}

\begin{table}[tbp]
\caption{Probabilities of rejection in Stage 1, Stage 2, combined probabilities of rejection, and the probabilities of continuing to Stage 2 for the misspecified model and for two boundary choices: O'Brien Fleming \& Pocock, with 200 cases, 200 controls; and 200 cases, 400 controls for $t=0.1, \lambda=1/2 \ \& \ \delta_0=0.165$. Here, we stop only for futility. }
    \centering
    \resizebox{\textwidth}{!}{
        \begin{tabular}{cccccccc}
            \hline
            Boundary & Case & $\mu_{f}$ & $\delta$ & $S1: P(Reject|H)$ & $S1: P(Continue|H)$ & $S2: P(Reject|H)$ & $P(Reject|H)$ \\ \midrule
          \multicolumn{8}{c}{200 cases and 200 controls} \\ \hline
            \multirow{3}{*}{O'Brien Fleming} & Null & $(1,1.1)$ & 0.165 & 0(0) & 0.488 (0.007) & 0.049 (0.003) & 0.049 (0.003) \\ 
            & \multirow{2}{*}{Alternate} & (1, 1.5) & 0.278 & 0(0) & 0.896 (0.004) & 0.544 (0.007) & 0.544 (0.007) \\ 
            & & (0.8, 2) & 0.467 & 0(0) & 0.999 (0.000) & 0.998 (0.001) & 0.998 (0.001) \\ \hline \multirow{3}{*}{Pocock} 
        & Null & $(1,1.1)$ & 0.165 & 0(0) & 0.279 (0.006) & 0.049 (0.003) & 0.049 (0.003) \\ 
        & \multirow{2}{*}{Alternate} 
        & (1, 1.5) & 0.278 & 0(0) & 0.735 (0.006) & 0.499 (0.007) & 0.499 (0.007)\\ 
        &  & (0.8, 2) & 0.467 & 0(0) & 0.996 (0.001) & 0.995 (0.001) & 0.995 (0.001)\\ \hline
        \multicolumn{8}{c}{200 cases and 400 controls} \\ \hline
        \multirow{3}{*}{O'Brien Fleming} 
        & Null & $(1,1.1)$ & 0.165 & 0(0) & 0.479 (0.007) & 0.045 (0.003) & 0.045 (0.003)  \\ 
        & \multirow{2}{*}{Alternate} 
        & (1, 1.5) & 0.278 & 0(0) & 0.936 (0.003) & 0.684 (0.007) & 0.684 (0.007) \\ 
        &  & (0.8, 2) & 0.467 & 0(0) & 1.000 (0.000) & 1.000 (0.000) & 1.000 (0.000) \\ \hline
        
        \multirow{3}{*}{Pocock} 
        & Null & $(1,1.1)$ & 0.165 & 0(0) & 0.277 (0.006) & 0.042 (0.003) & 0.042 (0.003) \\ 
        & \multirow{2}{*}{Alternate} 
        & (1, 1.5) & 0.278 & 0(0) & 0.837 (0.005) & 0.664 (0.007) & 0.664 (0.007) \\ 
        &  & (0.8, 2) & 0.467 & 0(0) & 1.000 (0.000) & 0.999 (0.000) & 0.999 (0.000) \\ \hline
    \end{tabular}
}
\label{tab:prob_0.1_fut_uneq}
\end{table}

\subsection{Group rotation}

We now evaluate the performance of the group rotation scheme discussed in Section \ref{subsec:rotation} in the context of testing the incremental performance of biomarkers. Suppose we consider $V=10$ sample units from each case and control.  We examine different sample sizes: 200 cases and 200 controls; 500 cases and 500 controls. For the control group, both the established marker and the new marker are generated from a normal distribution with mean 0 and variance 1. For the case group, the new marker follows a mixture of two normal distributions with variance 1 and mean 1.1 and 1.5, respectively, with $\gamma$ as the mixing parameter; the established marker follows a normal distribution with mean 1 and variance 1.  For the scenario where the linear logistic mode is correctly specified, the covariance between the established marker and each new marker is 0.2 for both cases and controls. For the scenario with the misspecified model, the covariance is 0.2 for the cases and 0.1 for the controls. Here $\gamma=\gamma_0$ implies that among cases, a proportion $\gamma_0$ of the new biomarkers are drawn from the normal distribution with mean 1.1, while the remaining new biomarkers are drawn from the normal distribution with mean 1.5.  Similar to the previous sections, we focus on the difference $ROC_f(0.1)-ROC_r(0.1)$ and test the following hypothesis: 

\begin{itemize}
    \item \textbf{Model correctly specified:} $  H_0: ROC_f(0.1)-ROC_r(0.1) \leq 0.141 \text{ vs } H_1: ROC_f(0.1)-ROC_r(0.1) > 0.141, $

    \item \textbf{Model misspecified:} $  H_0: ROC_f(0.1)-ROC_r(0.1) \leq 0.165 \text{ vs } H_1: ROC_f(0.1)-ROC_r(0.1) > 0.165, $
\end{itemize}
Here, the thresholds $\delta_0$ correspond to a null marker with mean 1.1 among cases, and a marker with mean 1.5 among cases is considered a useful marker with an incremental value of 0.259 and 0.278, respectively, among the correctly specified and misspecified models. Thus, $\gamma$ represents the proportion of null markers among all candidate markers evaluated using specimens from a biorepository.

\begin{figure}
    \centering
    \includegraphics[width=6cm , trim=0 0 4cm 0, clip]{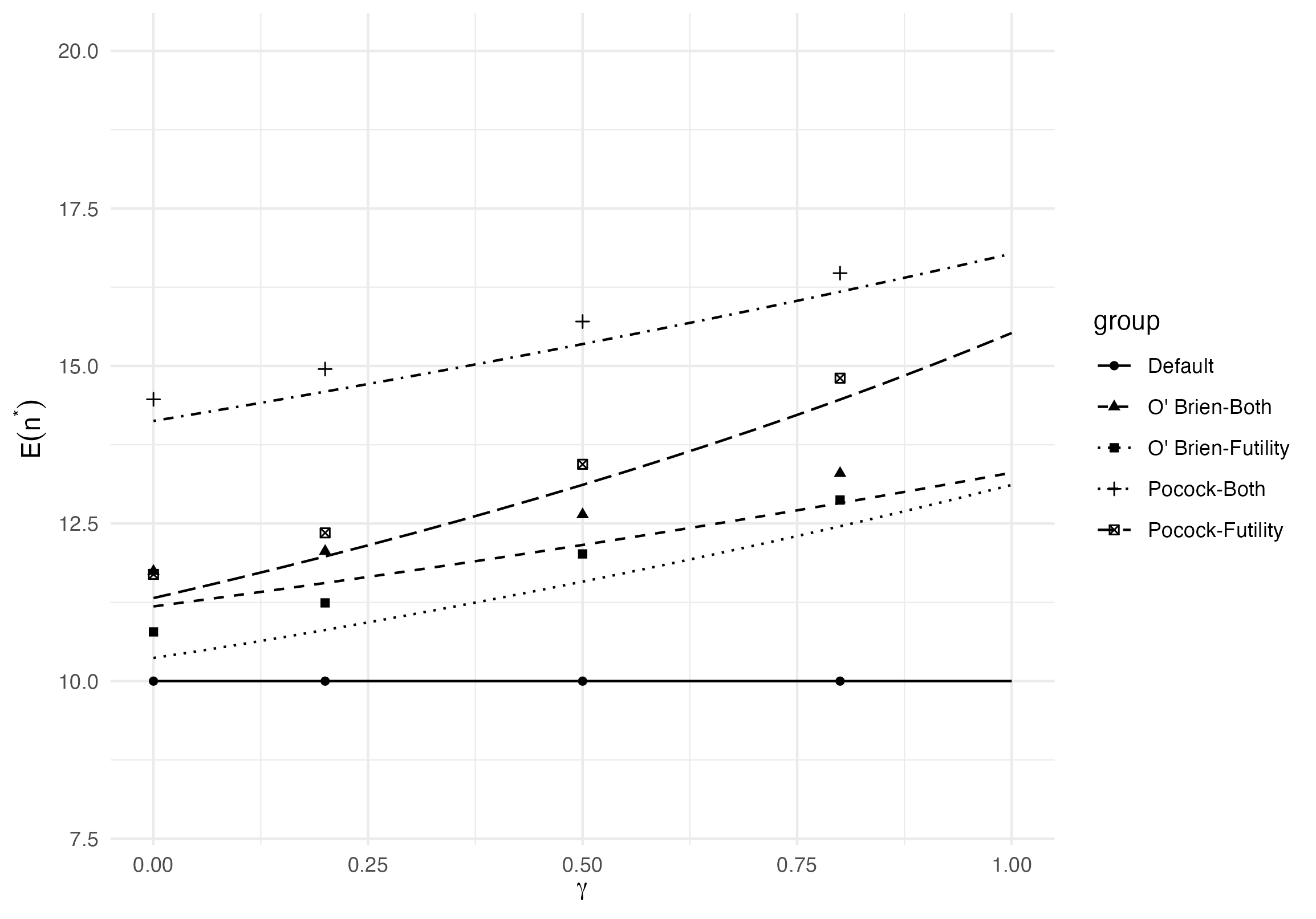}
\includegraphics[width=7.5cm]{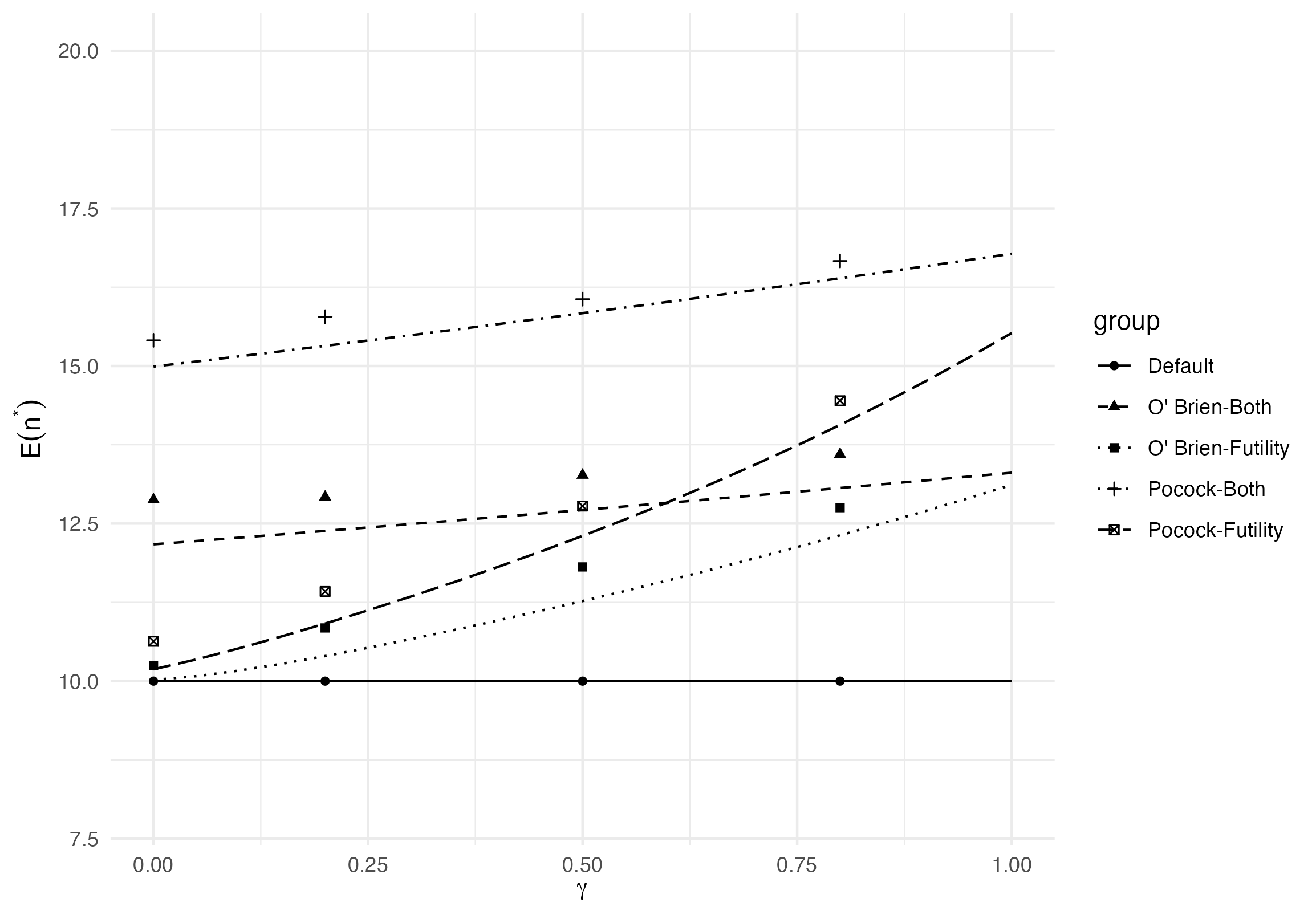}
 \includegraphics[width=6cm , trim=0 0 4cm 0, clip]{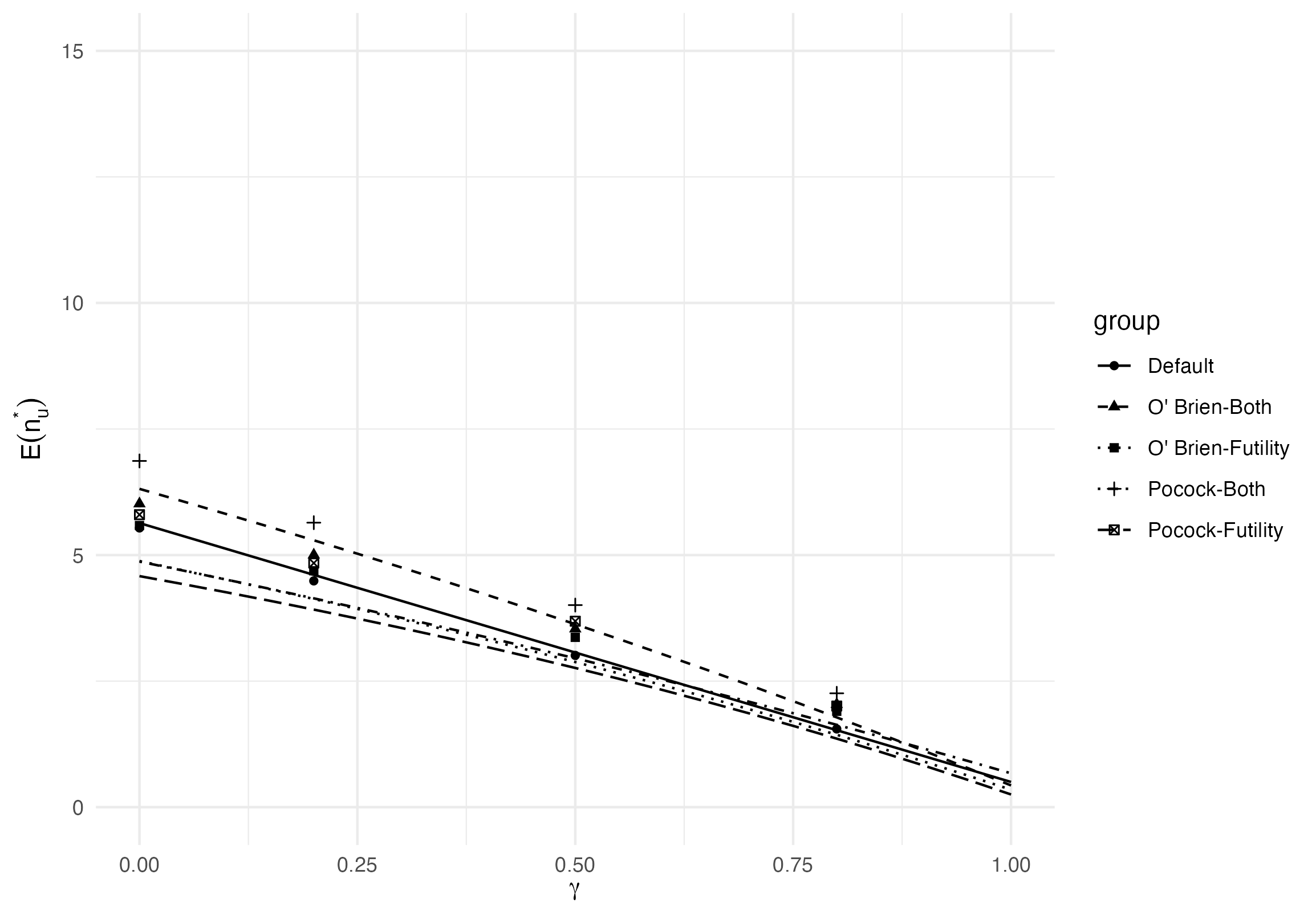}
\includegraphics[width=7.5cm]{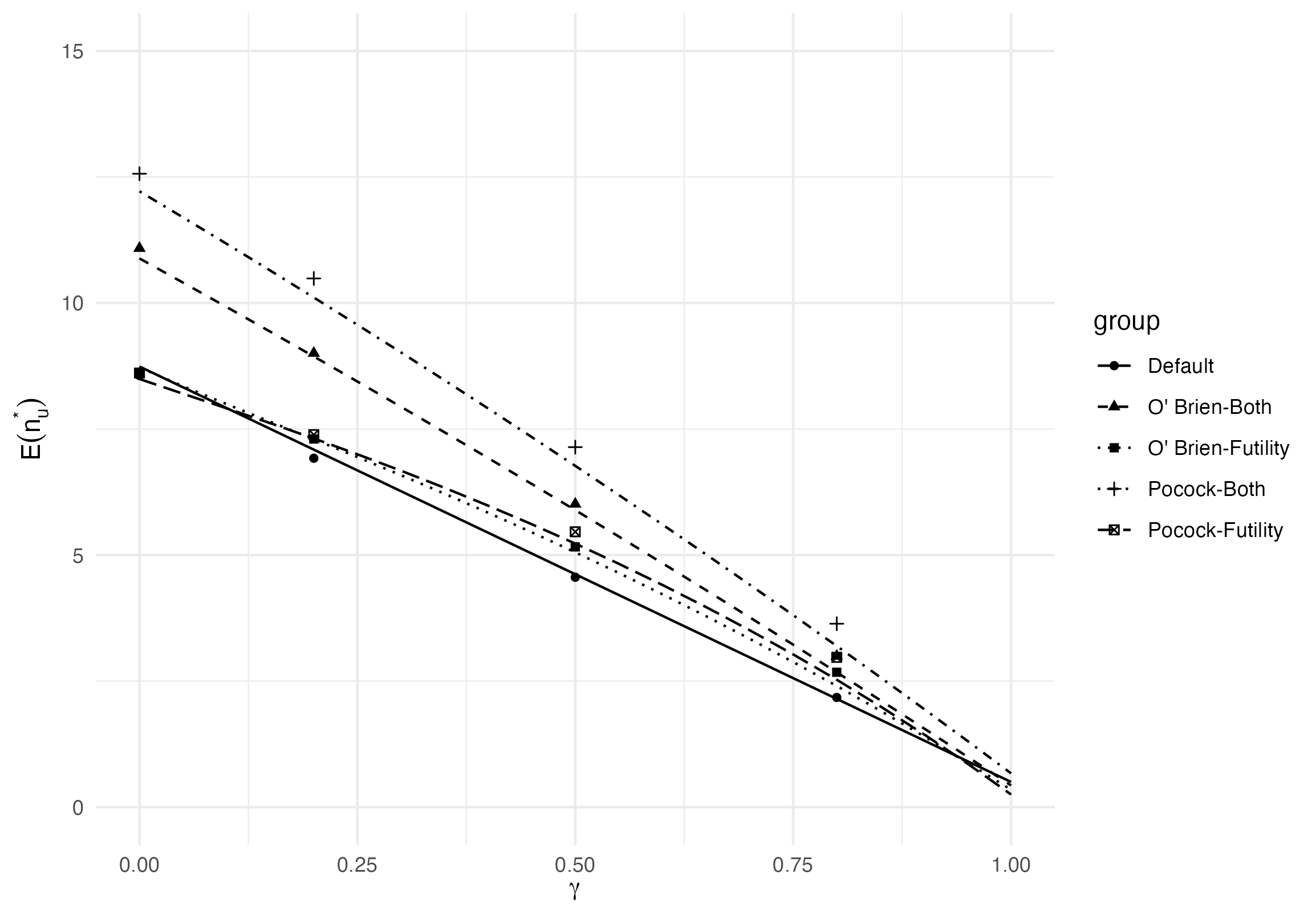}
  \includegraphics[width=6cm , trim=0 0 4cm 0, clip]{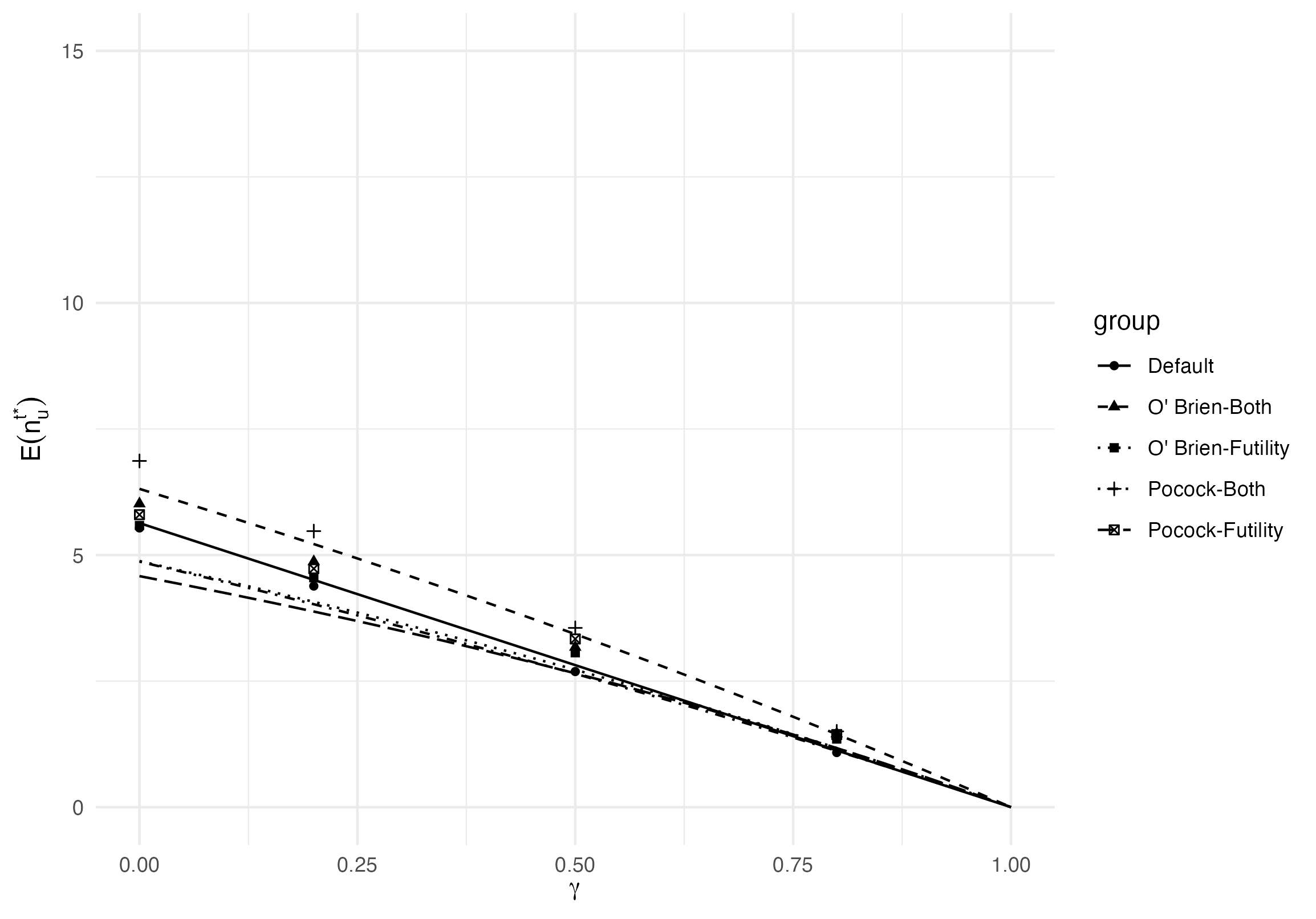}
     \includegraphics[width=7.5cm]{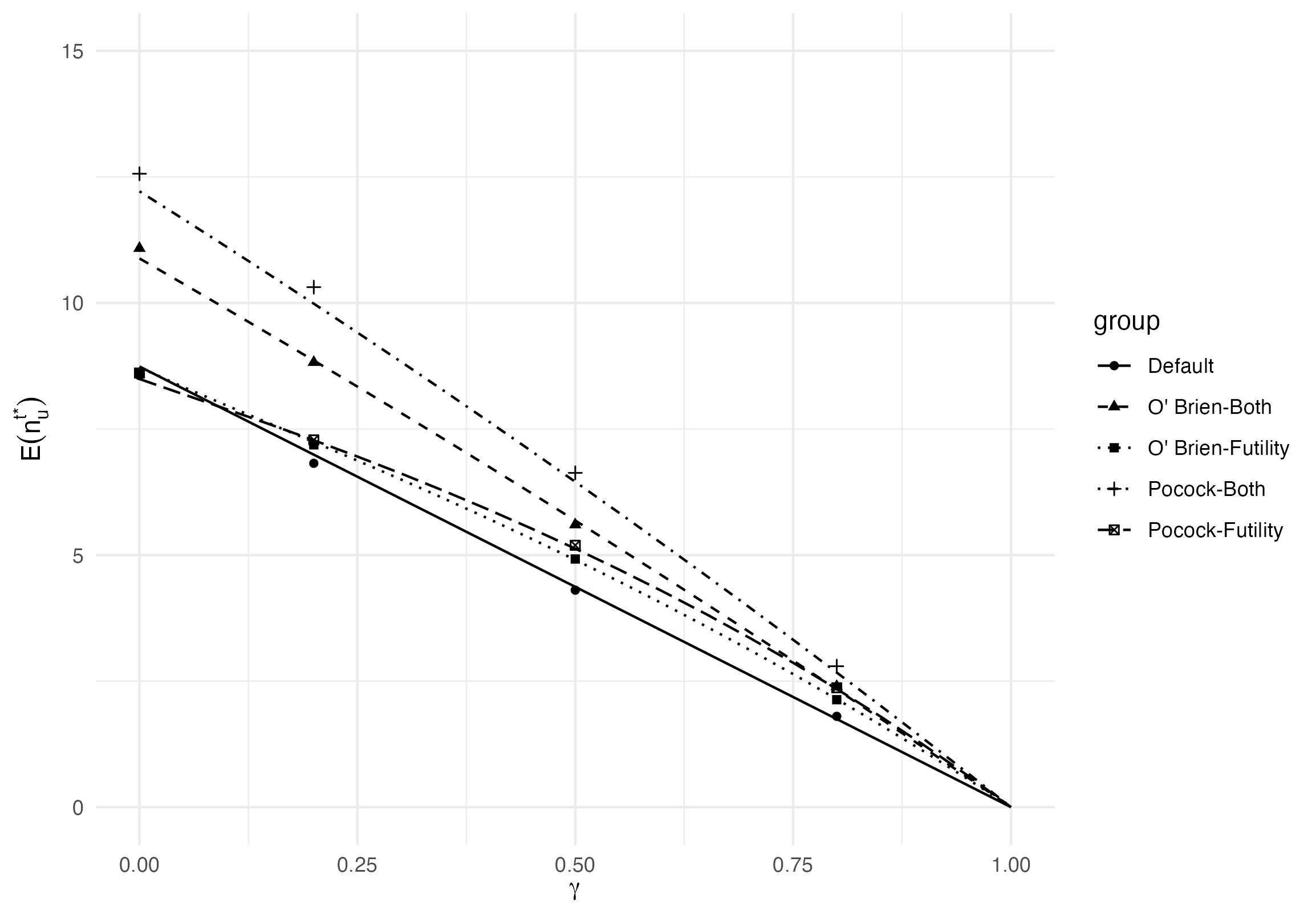}

    \caption{Comparison of the values of $E(n^*), E(n_u^*), E(n_u^{t*})$ evaluated computed based on the analytical formula and simulation for $t=0.1$ for the misspecified model. We consider two boundary criteria: O' Brien Fleming and Pocock and two stopping criteria: Stopping for both Futility and Efficacy and Stopping only for Futility. Results across combinations of these boundary and stopping criteria are compared with results from a non-sequential procedure, denoted as `Default'. The lines denote the values obtained from the analytical formula, while the dots in several shapes denote the simulation results at $\gamma=0, 0.2, 0.5, 0.8$. The figures to the left correspond to 200 cases and 200 controls, while the figures to the right correspond to 500 cases and 500 controls. }
    \label{fig:rotation_1_uneq}
\end{figure}

Figure \ref{fig:rotation_1_uneq} compares performance measures of the proposed testing strategy obtained from simulations with those based on the analytical expressions developed in Section \ref{subsec:rotation} for $t = 0.1$ under the misspecified model. Corresponding plots for the correctly specified model are provided in Figure \ref{fig:rotation_1_eq} of Appendix C.2 in the Supplementary Material, and similar plots for $t = 0.2$ under the misspecified model appear in Figure \ref{fig:rotation_1_0.2} of Appendix~C.4. Table \ref{tab:sim_rotation_500_uneq} presents a detailed comparison between analytical and simulation results for the misspecified model with 500 cases and 500 controls, including standard errors, for $t = 0.1$ and $\gamma \in \{0, 0.2, 0.5, 0.8\}$. A corresponding table for 200 cases and 200 controls is provided in Table \ref{tab:sim_rotation_200_uneq} of Appendix C.3, while Tables \ref{tab:sim_rotation_500_eq} and \ref{tab:sim_rotation_200_eq} containing results for the correctly specified model appear in Appendix C.2. All tables include results from a non-sequential procedure for reference, denoted as `Default'.

We note that, across all settings, the expected number of new biomarkers validated for incremental value over the established marker, $E(n^{*})$, is higher under the proposal design compared to the default design. This is expected since the default design always validates 10 markers using 10 volumes of samples. Our primary aim, however, is to see if our sequential procedure can improve $E(n_u^{t*})$ (the expected number of successfully validated biomarkers with truly significant incremental effects) compared to `Default'. Comparing the values across all tables and all figures, we see that for all choices of $\gamma$, we are successfully able to improve $E(n_u^{t*})$ compared to `Default' when stopping for both futility and efficacy. Stopping under futility slightly underperforms compared to the default design for smaller values of $\gamma$ ($\leq 0.20$), leading to smaller estimates of $E(n_u^*)$ (the expected number of biomarkers with the null hypothesis rejected) and $E(n_u^{t*})$. However, the performance of stopping under futility improves relative to the default design for higher $\gamma$. Across all settings, Pocock boundaries lead to the identification of a greater number of truly useful markers compared to O'Brien Fleming for each stopping criterion. Hence, Pocock is the preferred boundary condition. Comparing results between the two different sample sizes, it can be observed that a larger sample size leads to a larger value for $E(n_u^*)$ and $E(n_u^{t*})$, which is expected due to the increase in statistical power as the sample size increases. The values obtained from analytical formulations and simulations are reasonably close across both sample sizes and all values of $\gamma$, confirming the accuracy of our theoretical approximations. Importantly, this suggests that the analytical formula can be reliably used to design sample distribution schemes in future biomarker validation studies. By using the analytical formulas, researchers can efficiently plan specimen allocation and stopping rules without relying on extensive simulations.

\begin{table*}
  \caption{Comparison of the values of $E(n^*), E(n_u^*), E(n_u^{t*})$ computed based on the analytical formula and simulation for $t=0.1$ based on 500 cases and 500 controls for the misspecified model. We consider two boundary criteria: O' Brien Fleming and Pocock and two stopping criteria: Stopping for both Futility and Efficacy and Stopping only for Futility. Results across combinations of these boundary and stopping criteria are compared with results from a non-sequential procedure, denoted as `Default'. We consider four different values of the mixing parameter $\gamma=0, 0.2, 0.5, 0.8$. The values in brackets denote the corresponding standard errors based on 2000 simulations.}
    \centering
   \resizebox{\textwidth}{!}{  \begin{tabular}{ccccccccc}
    \hline
       Choice of $\gamma$ & Stop & & \multicolumn{2}{c}{$E(n^*)$ } & \multicolumn{2}{c}{$E(n_u^*)$} & \multicolumn{2}{c}{$E(n_u^{t*})$ } \\ \hline
    &  &  & True & Estimate  & True & Estimate  & True & Estimate \\ \midrule
\multirow{5}{*}{$\gamma=0$ } & \multirow{2}{*}{Both } & O' Brien Fleming & 12.171 & 12.878 (0.051) & 10.883 & 11.084 (0.076) & 10.883 & 11.084 (0.076) \\ 
& & Pocock & 14.990 & 15.408 (0.060) & 12.214 & 12.562 (0.089) & 12.214 & 12.562 (0.089) \\
& \multirow{2}{*}{Futility } & O' Brien Fleming & 10.018 & 10.244 (0.02) & 8.721 & 8.596 (0.05) & 8.721 & 8.596 (0.05) \\
& & Pocock & 10.187 & 10.632 (0.026) & 8.492 & 8.61 (0.048) & 8.492 & 8.61 (0.048) \\
& Default & & 10.000 & 10.000 (0.000) & 8.739 & 8.560 (0.046) & 8.739 & 8.560 (0.046) \\  \hline

\multirow{5}{*}{$\gamma=0.2$ } & \multirow{2}{*}{Both } & O' Brien Fleming & 12.383 & 12.922 (0.047) & 8.939 & 9.004 (0.082) & 8.858 & 8.823 (0.083) \\ 
& & Pocock & 15.319 & 15.781 (0.060) & 10.108 & 10.486 (0.094) & 9.986 & 10.313 (0.094) \\
& \multirow{2}{*}{Futility } & O' Brien Fleming & 10.398 & 10.844 (0.031) & 7.296 & 7.302 (0.064) & 7.241 & 7.186 (0.064)  \\
& & Pocock & 10.913 & 11.422 (0.038) & 7.313 & 7.388 (0.059) & 7.278 & 7.284 (0.06)  \\
& Default & & 10.000 & 10.000 (0.000) & 7.091 & 6.920 (0.067) & 6.991 & 6.820 (0.068)\\  \hline

\multirow{5}{*}{$\gamma=0.5$ } & \multirow{2}{*}{Both } & O' Brien Fleming & 12.715 & 13.267 (0.056) & 5.891 & 6.012 (0.079) & 5.685 & 5.602 (0.079) \\ 
& & Pocock & 12.715 & 16.060 (0.062) & 5.891 & 7.141 (0.092) & 5.685 & 6.631 (0.089)  \\
& \multirow{2}{*}{Futility } & O' Brien Fleming & 11.27 & 11.812 (0.04) & 5.054 & 5.162 (0.07) & 4.906 & 4.922 (0.069) \\
& & Pocock & 11.27 & 12.778 (0.051) & 5.054 & 5.462 (0.071) & 4.906 & 5.194 (0.071) \\
& Default & & 10.000 & 10.000 (0.000) & 4.619 & 4.560 (0.068) & 4.369 & 4.304 (0.067) \\  \hline

\multirow{5}{*}{$\gamma=0.8$ } & \multirow{2}{*}{Both } & O' Brien Fleming & 13.064 & 13.598 (0.056) & 2.675 & 2.996 (0.069) & 2.336 & 2.396 (0.063) \\ 
& & Pocock & 16.392 & 16.667 (0.063) & 3.196 & 3.639 (0.081) & 2.671 & 2.795 (0.070) \\
& \multirow{2}{*}{Futility } & O' Brien Fleming & 12.313 & 12.752 (0.047) & 2.404 & 2.676 (0.062) & 2.144 & 2.132 (0.058)  \\
& & Pocock & 14.065 & 14.448 (0.06) & 2.527 & 2.972 (0.063) & 2.345 & 2.368 (0.058) \\
& Default & & 10.000 & 10.000 (0.000) & 2.148 & 2.176 (0.057) & 1.748 & 1.802 (0.053)\\  \hline
    \end{tabular}}
    \label{tab:sim_rotation_500_uneq}
\end{table*}

\section{Data analysis}
\label{sec:data}
In this section, we demonstrate the application of the rotation scheme based on the proposed sequential testing procedure on the data from a multicenter EDRN pancreatic cancer reference set study \citep{haab2015definitive}. This data set consists of biomarker measurements from 98 cancer patients and 61 healthy individuals, and measurement of a number of biomarkers from all the participants. The biomarkers measured include CA19.9, the established marker for pancreatic cancer diagnosis, and multiple new candidate markers from several laboratories including MUC5AC, GDF15, TFPI, TNC.FIII.C, AT, ATQ, ATATQ,                   TIMP1, LRG1, APOA2.AT, APOA2.TQ, and APOA2.i.Index. 

Our objective is to assess whether the proposed rotation scheme can enhance the validation of useful biomarker combinations compared to a non-sequential design. To achieve this, we will first identify 
potentially useful biomarkers that could improve pancreatic cancer detection when combined with CA 19.9 in our dataset. In this context, we refer to the recent works by \citet{honda2015plasma} and \citet{kashiro2024clinical} who have established the serum-based ATATQ and plasma-based APOA2.i.Index as potential biomarkers for early-stage pancreatic cancer detection. A straightforward application of logistic regression on the dataset, regressing case/control status on the log-transformed CA 19.9 and ATATQ values, yields a highly significant p-value for  ATATQ, signifying its importance in pancreatic cancer detection when combined with CA 19.9. Similar results follow for APOA2.i.Index. Therefore, these two markers, in combination with CA19.9, present themselves as valuable biomarkers for pancreatic cancer detection.  

With the objective of identifying new markers that can potentially be combined with CA19.9 to be used in pancreatic cancer surveillance, we consider testing of incremental value in sensitivity at 90\% specificity with the following hypothesis: \begin{equation*}
       H_0: ROC_f(0.1)-ROC_r(0.1) \leq 0.01 \text{ vs } H_1: ROC_f(0.1)-ROC_r(0.1) > 0.01. 
\end{equation*} 
 
Based on this hypothesis test, we will verify whether our proposed approach helps in saving specimens when validating biomarkers from a pool of candidates, similar to those in the pancreatic cancer reference study. We proceed by estimating the number of markers that can be evaluated, the number of significant incremental values identified,  and the number of truly useful incremental performances successfully validated, that is, we wish to estimate the quantities $E(n^*), E(n_u^*), E(n_u^{t*})$. 

To estimate $E(n^*), E(n_u^*)$ and $E(n_u^{t*})$, we generate 1000 bootstrap samples from the original data, stratified on cases and controls, and consider 50 volumes of specimens for each bootstrap sample, which are utilized to measure 50 markers selected with replacement from the pool of markers available. We choose $\lambda= 1/2$ and $1/3$, that is, we consider two scenarios with the first stage sample half or one-third of the total number of individuals in the study. The estimates and the corresponding standard errors for $t=0.1$ are noted in Table \ref{tab:real_rotation}. We also note the default scenario with a non-sequential testing procedure in the row corresponding to `Default'.  We observe that, compared to the default procedure, the sequential procedure can evaluate a larger number of biomarkers, including more truly useful markers. For example, while stopping for both futility and efficacy with Pocock boundaries and $\lambda = 1/3$, the expected number of biomarkers evaluated increases from 50 to 95.48, an improvement of approximately 91\% compared to `Default'. Similarly, the expected number of biomarkers with rejected null hypotheses rises from 2.16 to 5.91 (a 174\% increase), and the expected number of truly useful biomarkers validated grows from 1.40 to 3.30 (a 136\% increase).  Across all scenarios, using the Pocock boundary results in validating more markers than the O'Brien Fleming boundary. Specifically, with the Pocock boundary, a smaller $\lambda$ leads to evaluating more biomarkers, reflected in higher values of $E(n^*)$, $E(n_u^*)$, and $E(n_u^{t*})$ while the scenario reverses for the O'Brien Fleming boundary, where a bigger $\lambda$ works better. Stopping for futility alone is less effective than stopping for both futility and efficacy, but still shows modest improvement over the default design. These improvements in the number of useful biomarkers successfully validated using the proposed group-sequential rotation strategy highlight the practical advantage of our proposed methodology for biorepository-based studies with limited specimens.


\begin{table}[tbp]
  \caption{Estimates of $E(n^*), E(n_u^*), E(n_u^{t*})$ based on 1000 bootstrap samples for $t=0.1 \ \ \& \ \ \delta_0=0.01$ with O'Brien Fleming and Pocock boundaries, and for $\lambda=1/2$ and $1/3$. We consider two stopping criteria: stopping for futility \& efficacy, and stopping for futility only.}
    \centering
   \resizebox{\textwidth}{!}{   \begin{tabular}{cccccccc}
    \hline
    && \multicolumn{3}{c}{O' Brien Fleming} & \multicolumn{3}{c}{Pocock} \\ \hline
      Stop & $\lambda$  &  $E(n^*)$ & $E(n_u^*)$ & $E(n_u^{t*})$ &  $E(n^*)$ & $E(n_u^*)$ & $E(n_u^{t*})$\\ \midrule
      Default &  & 50.00 (0.00) & 2.16 (0.14)&  1.40 (0.09) & 50.00 (0.00) & 2.16 (0.14)&  1.40 (0.09)\\ \hline
      Both & 1/2 & 68.67 (0.16)& 3.67 (0.17)& 2.32 (0.10) & 89.28 (0.19)& 4.92 (0.17) & 3.01 (0.11)  \\ \hline
      Futility & 1/2 & 65.21 (0.14) & 2.22  (0.13) & 1.57 (0.09) & 83.91 (0.26) & 2.63(0.16) & 1.82 (0.11) \\ \hline
      Both & 1/3 & 51.83 (0.04) & 2.66 (0.15)&  1.53 (0.08) &95.48 (0.37) & 5.91 (0.18) &  3.30 (0.11)\\ \hline
      Futility & 1/3 & 51.47 (0.04)  & 2.34 (0.15) & 1.39 (0.08) & 82.56 (0.34) & 2.93 (0.17) &  1.98 (0.12) \\ \hline
    \end{tabular}}
    \label{tab:real_rotation}
\end{table}


\section{Discussion}
\label{sec: disc}
This work proposes a two-stage group sequential hypothesis testing approach in a case-control study to evaluate the enhancement in disease detection performance when a new marker is incorporated into a panel of established biomarkers. The sequential procedure is advantageous in situations with limited specimen availability, enabling an optimal allocation of biomarker measurements to maximize the number of biomarkers evaluated. We derive the asymptotic distribution of the change in sensitivity at a specified level of specificity using a logistic regression working model and leverage these asymptotic properties to formulate the test statistic for the testing procedure. Further, we explore a rotation strategy following \citet{wang2021strategies}, which makes use of this testing procedure to accommodate a large number of biomarkers for validation.

While this paper focuses on testing incremental value through comparison of two working models that may be misspecified, the asymptotic framework we develop applies more broadly to evaluating biomarker panel performance using working models—an issue that frequently arises in biomarker research. To illustrate this broader applicability, we have included a small simulation study in Appendix C.5 of the Supplementary Material, where we have computed type I error and power while conducting sequential testing for a single biomarker panel (rather than incremental value). The results confirm that our framework maintains nominal type I error and achieves high power as the effect size increases.

As has been mentioned before, this paper deals with testing procedures for the incremental effects of a new biomarker combined with a panel of established biomarkers. This also applies to settings when one is interested in identifying new biomarkers with significant incremental value over clinical covariates known to be risk predictors for diseases. A more comprehensive testing procedure could incorporate a set of evaluations to determine whether a new biomarker has significant performance itself and whether this new biomarker combined with established biomarkers has significant incremental performance. Also, in settings that implement early stopping for futility, the proposed sequential procedure can be used to develop Uniformly Minimum Variance Unbiased Estimators (UMVUE) of the incremental value among markers that are successfully validated. We leave these areas as topics for future research.

\section*{Acknowledgments}
 This work was supported by NIH grants NCI R01 CA277133 and U24 CA086368.

\bibliographystyle{plainnat}  
\bibliography{ref}            

\begin{thebibliography}{27}
\providecommand{\natexlab}[1]{#1}
\providecommand{\url}[1]{\texttt{#1}}
\expandafter\ifx\csname urlstyle\endcsname\relax
  \providecommand{\doi}[1]{doi: #1}\else
  \providecommand{\doi}{doi: \begingroup \urlstyle{rm}\Url}\fi

\bibitem[Chen et~al.(2013)Chen, Samuelson, Gallas, Kang, Sahiner, and Petrick]{chen2013assessment}
Weijie Chen, Frank~W Samuelson, Brandon~D Gallas, Le~Kang, Berkman Sahiner, and Nicholas Petrick.
\newblock On the assessment of the added value of new predictive biomarkers.
\newblock \emph{BMC medical research methodology}, 13:\penalty0 1--9, 2013.

\bibitem[Cook(2018)]{cook2018quantifying}
Nancy~R Cook.
\newblock Quantifying the added value of new biomarkers: how and how not.
\newblock \emph{Diagnostic and Prognostic Research}, 2\penalty0 (1):\penalty0 14, 2018.

\bibitem[Emerson and Fleming(1989)]{emerson1989symmetric}
Scott~S Emerson and Thomas~R Fleming.
\newblock Symmetric group sequential test designs.
\newblock \emph{Biometrics}, pages 905--923, 1989.

\bibitem[Feng et~al.(2013)Feng, Kagan, Pepe, Thornquist, Ann~Rinaudo, Dahlgren, Krueger, Zheng, Patriotis, Huang, et~al.]{feng2013early}
Ziding Feng, Jacob Kagan, Margaret Pepe, Mark Thornquist, Jo~Ann~Rinaudo, Jackie Dahlgren, Karl Krueger, Yingye Zheng, Christos Patriotis, Ying Huang, et~al.
\newblock The early detection research network's specimen reference sets: paving the way for rapid evaluation of potential biomarkers.
\newblock \emph{Clinical chemistry}, 59\penalty0 (1):\penalty0 68--74, 2013.

\bibitem[Gordon~Lan and DeMets(1983)]{gordon1983discrete}
KK~Gordon~Lan and David~L DeMets.
\newblock Discrete sequential boundaries for clinical trials.
\newblock \emph{Biometrika}, 70\penalty0 (3):\penalty0 659--663, 1983.

\bibitem[Haab et~al.(2015)Haab, Huang, Balasenthil, Partyka, Tang, Anderson, Allen, Sasson, Zeh, Kaul, et~al.]{haab2015definitive}
Brian~B Haab, Ying Huang, Seetharaman Balasenthil, Katie Partyka, Huiyuan Tang, Michelle Anderson, Peter Allen, Aaron Sasson, Herbert Zeh, Karen Kaul, et~al.
\newblock Definitive characterization of ca 19-9 in resectable pancreatic cancer using a reference set of serum and plasma specimens.
\newblock \emph{PloS one}, 10\penalty0 (10):\penalty0 e0139049, 2015.

\bibitem[Honda et~al.(2015)Honda, Kobayashi, Okusaka, Rinaudo, Huang, Marsh, Sanada, Sasajima, Nakamori, Shimahara, et~al.]{honda2015plasma}
Kazufumi Honda, Michimoto Kobayashi, Takuji Okusaka, Jo~Ann Rinaudo, Ying Huang, Tracey Marsh, Mitsuaki Sanada, Yoshiyuki Sasajima, Shoji Nakamori, Masashi Shimahara, et~al.
\newblock Plasma biomarker for detection of early stage pancreatic cancer and risk factors for pancreatic malignancy using antibodies for apolipoprotein-aii isoforms.
\newblock \emph{Scientific reports}, 5\penalty0 (1):\penalty0 15921, 2015.

\bibitem[Jennison and Turnbull(1999)]{jennison1999group}
Christopher Jennison and Bruce~W Turnbull.
\newblock \emph{Group sequential methods with applications to clinical trials}.
\newblock CRC Press, 1999.

\bibitem[Kashiro et~al.(2024)Kashiro, Kobayashi, Oh, Miyamoto, Atsumi, Nagashima, Takeuchi, Nara, Hijioka, Morizane, et~al.]{kashiro2024clinical}
Ayumi Kashiro, Michimoto Kobayashi, Takanori Oh, Mitsuko Miyamoto, Jun Atsumi, Kengo Nagashima, Keiko Takeuchi, Satoshi Nara, Susumu Hijioka, Chigusa Morizane, et~al.
\newblock Clinical development of a blood biomarker using apolipoprotein-a2 isoforms for early detection of pancreatic cancer.
\newblock \emph{Journal of Gastroenterology}, 59\penalty0 (3):\penalty0 263--278, 2024.

\bibitem[Koopmeiners et~al.(2012)Koopmeiners, Feng, and Pepe]{koopmeiners2012conditional}
Joseph~S Koopmeiners, Ziding Feng, and Margaret~Sullivan Pepe.
\newblock Conditional estimation after a two-stage diagnostic biomarker study that allows early termination for futility.
\newblock \emph{Statistics in Medicine}, 31\penalty0 (5):\penalty0 420--435, 2012.

\bibitem[Manski and Lerman(1977)]{manski1977estimation}
Charles~F Manski and Steven~R Lerman.
\newblock The estimation of choice probabilities from choice based samples.
\newblock \emph{Econometrica: Journal of the Econometric Society}, pages 1977--1988, 1977.

\bibitem[McIntosh and Pepe(2002)]{mcintosh2002combining}
Martin~W McIntosh and Margaret~Sullivan Pepe.
\newblock Combining several screening tests: optimality of the risk score.
\newblock \emph{Biometrics}, 58\penalty0 (3):\penalty0 657--664, 2002.

\bibitem[O'Brien and Fleming(1979)]{o1979multiple}
Peter~C O'Brien and Thomas~R Fleming.
\newblock A multiple testing procedure for clinical trials.
\newblock \emph{Biometrics}, pages 549--556, 1979.

\bibitem[Park et~al.(2021)Park, Yuan, Ning, Liu, and Feng]{park2021hybrid}
Yeonhee Park, Ying Yuan, Jing Ning, Suyu Liu, and Ziding Feng.
\newblock Hybrid design evaluating new biomarkers when there is an existing screening test.
\newblock \emph{Statistics in medicine}, 40\penalty0 (8):\penalty0 2037--2054, 2021.

\bibitem[Pepe et~al.(2001)Pepe, Etzioni, Feng, Potter, Thompson, Thornquist, Winget, and Yasui]{pepe2001phases}
Margaret~Sullivan Pepe, Ruth Etzioni, Ziding Feng, John~D Potter, Mary~Lou Thompson, Mark Thornquist, Marcy Winget, and Yutaka Yasui.
\newblock Phases of biomarker development for early detection of cancer.
\newblock \emph{Journal of the National Cancer Institute}, 93\penalty0 (14):\penalty0 1054--1061, 2001.

\bibitem[Pepe et~al.(2009)Pepe, Feng, Longton, and Koopmeiners]{pepe2009conditional}
Margaret~Sullivan Pepe, Ziding Feng, Gary Longton, and Joseph Koopmeiners.
\newblock Conditional estimation of sensitivity and specificity from a phase 2 biomarker study allowing early termination for futility.
\newblock \emph{Statistics in Medicine}, 28\penalty0 (5):\penalty0 762--779, 2009.

\bibitem[Pocock(1977)]{pocock1977group}
Stuart~J Pocock.
\newblock Group sequential methods in the design and analysis of clinical trials.
\newblock \emph{Biometrika}, 64\penalty0 (2):\penalty0 191--199, 1977.

\bibitem[Prentice and Pyke(1979)]{prentice1979logistic}
Ross~L Prentice and Ronald Pyke.
\newblock Logistic disease incidence models and case-control studies.
\newblock \emph{Biometrika}, 66\penalty0 (3):\penalty0 403--411, 1979.

\bibitem[Qin and Zhang(2003)]{qin2003using}
Jing Qin and Biao Zhang.
\newblock Using logistic regression procedures for estimating receiver operating characteristic curves.
\newblock \emph{Biometrika}, 90\penalty0 (3):\penalty0 585--596, 2003.

\bibitem[Qin and Zhang(2010)]{qin2010best}
Jing Qin and Biao Zhang.
\newblock Best combination of multiple diagnostic tests for screening purposes.
\newblock \emph{Statistics in medicine}, 29\penalty0 (28):\penalty0 2905--2919, 2010.

\bibitem[Tamhane et~al.(2021)Tamhane, Xi, and Gou]{tamhane2021group}
Ajit~C Tamhane, Dong Xi, and Jiangtao Gou.
\newblock Group sequential holm and hochberg procedures.
\newblock \emph{Statistics in Medicine}, 40\penalty0 (24):\penalty0 5333--5350, 2021.

\bibitem[Tang et~al.(2008)Tang, Emerson, and Zhou]{tang2008nonparametric}
Liansheng Tang, Scott~S Emerson, and Xiao-Hua Zhou.
\newblock Nonparametric and semiparametric group sequential methods for comparing accuracy of diagnostic tests.
\newblock \emph{Biometrics}, 64\penalty0 (4):\penalty0 1137--1145, 2008.

\bibitem[Tayob et~al.(2016)Tayob, Do, and Feng]{tayob2016unbiased}
Nabihah Tayob, Kim-Anh Do, and Ziding Feng.
\newblock Unbiased estimation of biomarker panel performance when combining training and testing data in a group sequential design.
\newblock \emph{Biometrics}, 72\penalty0 (3):\penalty0 888--896, 2016.

\bibitem[Wang et~al.(2021)Wang, Huang, and Feng]{wang2021strategies}
Lu~Wang, Ying Huang, and Ziding Feng.
\newblock Strategies for validating biomarkers using data from a reference set.
\newblock \emph{Biostatistics}, 22\penalty0 (2):\penalty0 298--314, 2021.

\bibitem[Wassmer and Brannath(2016)]{wassmer2016group}
Gernot Wassmer and Werner Brannath.
\newblock \emph{Group sequential and confirmatory adaptive designs in clinical trials}, volume 301.
\newblock Springer, 2016.

\bibitem[Xie and Manski(1989)]{xie1989logit}
Yu~Xie and Charles~F Manski.
\newblock The logit model and response-based samples.
\newblock \emph{Sociological Methods \& Research}, 17\penalty0 (3):\penalty0 283--302, 1989.

\bibitem[Zhao et~al.(2015)Zhao, Zheng, Prentice, and Feng]{zhao2015two}
Shanshan Zhao, Yingye Zheng, Ross~L Prentice, and Ziding Feng.
\newblock Two-stage biomarker panel study and estimation allowing early termination for futility.
\newblock \emph{Biostatistics}, 16\penalty0 (4):\penalty0 799--812, 2015.

\end{thebibliography}
\appendix
\section*{Supplementary Material}
\section{Proof of Theorem 1:}
We focus on $\widehat{ROC}_f(t)-ROC_f(t)$ and look at its asymptotics. 
\begin{align*}
    &\hspace{-25mm}\widehat{ROC}_f(t)-ROC_f(t) \\ 
    &\hspace{-20mm}= [1-\widehat{F}_{D_1, f}(\hat{\beta}_f,\widehat{F}_{D_0, f}^{-1}(\hat{\beta}_f,1-t))]- [1-{F}_{D_1, f}(\hat{\beta}_f,\widehat{F}_{D_0, f}^{-1}(\hat{\beta}_f,1-t))] \\ 
    &\hspace{-20mm} +([1-{F}_{D_1, f}(\hat{\beta}_f,\widehat{F}_{D_0, f}^{-1}(\hat{\beta}_f,1-t)]-[1-{F}_{D_1, f}(\beta_f^*,\widehat{F}_{D_0, f}^{-1}(\hat{\beta}_f,1-t)]) \\ 
    &\hspace{-20mm} + ([1-{F}_{D_1, f}({\beta}_f^*,\widehat{F}_{D_0, f}^{-1}(\hat{\beta}_f,1-t)]-[1-{F}_{D_1, f}({\beta}_f^*,{F}_{D_0, f}^{-1}(\hat{\beta}_f,1-t)]) \\ 
    &\hspace{-20mm} + ([1-{F}_{D_1, f}({\beta}_f^*,{F}_{D_0, f}^{-1}(\hat{\beta}_f,1-t)]-[1-{F}_{D_1, f}({\beta}_f^*,{F}_{D_0, f}^{-1}({\beta}_f^*,1-t)]) \\ 
    &\hspace{-20mm}= A_1+A_2+A_3+A_4
\end{align*} where 
\begin{align*}
    A_1 &= (1-\widehat{F}_{D_1, f}(\hat{\beta}_f,\widehat{F}_{D_0, f}^{-1}(\hat{\beta}_f,1-t)))- (1-{F}_{D_1, f}(\hat{\beta}_f,\widehat{F}_{D_0, f}^{-1}(\hat{\beta}_f,1-t))) \\
    A_2 &= (1-{F}_{D_1, f}(\hat{\beta}_f,\widehat{F}_{D_0, f}^{-1}(\hat{\beta}_f,1-t)))-(1-{F}_{D_1, f}({\beta}_f^*,\widehat{F}_{D_0, f}^{-1}(\hat{\beta}_f,1-t))) \\ 
    A_3 &= (1-{F}_{D_1, f}({\beta}_f^*,\widehat{F}_{D_0, f}^{-1}(\hat{\beta}_f,1-t)))-(1-{F}_{D_1, f}({\beta}_f^*,{F}_{D_0, f}^{-1}(\hat{\beta}_f,1-t))) \\ 
    A_4 &= (1-{F}_{D_1, f}({\beta}_f^*,{F}_{D_0, f}^{-1}(\hat{\beta}_f,1-t)))-(1-{F}_{D_1, f}({\beta}_f^*,{F}_{D_0, f}^{-1}({\beta}_f^*,1-t))) 
\end{align*}

Let $n_1=\lambda N_1$, $n_0=\lambda N_0$ where $n_1$ and $n_0$ denote the number of cases and controls selected in Stage 1, with $n=n_1+n_0$. We denote $1-\widehat{F}$ by $\widehat{S}$ from now on. Let us look at the distribution of $\sqrt{n_1}A_1$. Also, let $u={F}_{D_0, f}^{-1}({\beta}_f,1-t)$.  


Let us look at the terms $A_1$, $A_2$, $A_3$, $A_4$ one by one.
 \noindent \begin{align*}
A_1 &=  (1-\widehat{F}_{D_1, f}(\hat{\beta}_f,\widehat{F}_{D_0, f}^{-1}(\hat{\beta}_f,1-t)))- (1-{F}_{D_1, f}(\hat{\beta}_f,\widehat{F}_{D_0, f}^{-1}(\hat{\beta}_f,1-t))) \\ &= (\widehat{S}_{D_1, f}(\beta_f^*,u)-{S}_{D_1, f}(\beta_f^*,u))+o_p(n^{-1/2}) \text{ (By equicontinuity) } \\ &=  \frac{1}{n_1} \sum_{i:D_i=1} \left(I(X_i^{f'}\beta_f^* >u) - P(X^{f'}\beta_f^* >u)  \right) + o_p(n^{-1/2})  \label{eq:a1} \\ &\approx \frac{1}{n_1} \sum_{i:D_i=1} \left(I(X_i^{f'}\beta_f^* >u) - P(X^{f'}\beta_f^* >u)  \right)
\end{align*}

\begin{align*}
A_2 &= {S}_{D_1, f}(\hat{\beta}_f,u) -{S}_{D_1, f}({\beta}_f^*,u)  
= \frac{\partial}{\partial \beta_f^*} ({S}_{D_1, f}(\beta_f^*,u))'(\hat{\beta}_f-\beta_f^*) \\ &\approx \frac{\partial}{\partial \beta_f^*} ({S}_{D_1, f}(\beta_f^*,u))'E(-\frac{\partial^2 l(\beta_f^*)}{\partial \beta_f^{*2}})^{-1} U(\beta_f^*)
\end{align*} where $U(\beta_f^*)$ denotes the derivative of the log-likelihood $l(\alpha,\beta_f^*)$ with respect to $\beta_f^*$. For a logistic regression working model, we have the log-likelihood function given by \begin{equation*}
    l(\alpha,\beta_f^*) = \sum_{i=1}^n\left(D_i (\alpha+X_i^{f'}\beta_f^*) - \log(1+ exp(\alpha+ X_i^{f'}\beta_f^*)) \right)
\end{equation*} Thus, we get, \begin{equation*}
   U(\beta_f^*) = \sum_{i=1}^n X_i^f(D_i-p_i) 
\end{equation*} where $p_i= \frac{1}{1+exp(-(\alpha+X_i^{f'}\beta_f^*))}$ and \begin{equation*}
    \frac{\partial^2 l(\alpha,\beta_f^*)}{\partial \beta_f^{*2}} = -\sum_{i=1}^n X_i^f X_i^{f'}p_i (1-p_i)
\end{equation*} Hence, we get, \begin{equation*}
    A_2 = \sum_{i=1}^n \left(\frac{\partial}{\partial \beta_f^*} ({S}_{D_1, f}(\beta_f^*,u))' E \left(\sum_{i=1}^n X_i^f X_i^{f'}p_i (1-p_i) \right)\right)X_i^f(D_i-p_i) 
\end{equation*}
For ease of notations, let us denote $g= \frac{\partial}{\partial \beta_f^*} ({S}_{D_1, f}(\beta_f^*,u))$, $E_1= E \left(\sum_{i=1}^n X_i^f X_i^{f'}p_i (1-p_i) \right)$. Thus, $$A_2= \sum_{i=1}^n g' E_1 X_i^f(D_i-p_i)$$

Next, \begin{equation*}
    A_3 = (1-{F}_{D_1, f}(\beta_f^*,\widehat{F}_{D_0, f}^{-1}(\hat{\beta}_f,1-t)))-(1-{F}_{D_1, f}(\beta_f^*,{F}_{D_0, f}^{-1}(\hat{\beta}_f,1-t))) 
\end{equation*} Now,
\begin{align*}
      \widehat{F}_{D_0,f}^{-1}({\hat{\beta}}_f,1-t)-{F}_{D_0,f}^{-1}({\hat{\beta}}_f,1-t) 
    &= \widehat{F}_{D_0,f}^{-1}({{\beta}}_f^*,1-t)-{F}_{D_0,f}^{-1}({{\beta}}_f^*,1-t) + o_p(n_0^{-1/2}) \\ &=- \frac{\widehat{F}_{D_0,f}({{\beta}}_f^*,u)-(1-t)}{{f}_{D_0,f}({{\beta}}_f^*,u)} + o_p(n_0^{-1/2}) \text{ (Using Bahadur representation) } \\ &= - \frac{1}{n_0 {f}_{D_0,f}({{\beta}}_f^*,u)} \sum_{i:D_i=0} \left(I(X_i^{f'}\beta_f^*\leq u)- (1-t)\right) + o_p(n_0^{-1/2})
\end{align*} Hence, we get, \begin{equation*}
     A_3 \approx  \frac{{f}_{D_1,f}({{\beta}}_f^*,u)}{n_0 {f}_{D_0,f}({{\beta}}_f^*,u)} \sum_{i:D_i=0} \left(I(X_i^{f'}\beta_f^*\leq u)- (1-t)\right)
\end{equation*}

Similar to $A_2$, $A_4$ can be written as \begin{equation*}
    A_4 = \sum_{i=1}^n (-{f}_{D_1,f}({{\beta}}_f^*,u)\frac{\partial}{\partial \beta_f^*} ({F}_{D_0, f}^{-1}(\beta_f^*,1-t))  E \left(\sum_{i=1}^n X_i^f X_i^{f'}p_i (1-p_i) \right)X_i^f(D_i-p_i) 
\end{equation*} Let us denote $h= -{f}_{D_1,f}({{\beta}}_f^*,u)\frac{\partial}{\partial \beta_f^*} ({F}_{D_0, f}^{-1}(\beta_f^*,1-t))$. Then, $A_4$ can be written as $$A_4= \sum_{i=1}^n h'E_1 X_i^f(D_i-p_i)$$ Hence, we can write \begin{equation*}
  \sqrt{n}  \begin{pmatrix}A_1 \\ A_2 \\ A_3 \\ A_4 \end{pmatrix} = \sqrt{n} \frac{1}{n} \sum_{i=1}^n  \begin{pmatrix}\frac{I(D_i=1)\left(I(X_i^{f'}\beta_f^* >u) - P(X^{f'}\beta_f^* >u) \right)}{\pi_{case}} 
  \\ g' E_1 X_i^f(D_i-p_i)  
  \\ \frac{I(D_i=0)\left(I(X_i^{f'}\beta_f^*\leq u)- (1-t)\right)}{\pi_0} 
  \\ h'  E_1 X_i^f(D_i-p_i) \end{pmatrix}
\end{equation*} where the vectors $\begin{pmatrix}\frac{I(D_i=1)\left(I(X_i^{f'}\beta_f^* >u) - P(X^{f'}\beta_f^* >u) \right)}{\pi_{case}} 
  \\ g' E_1 X_i^f(D_i-p_i)  
  \\ \frac{I(D_i=0)\left(I(X_i^{f'}\beta_f^*\leq u)- (1-t)\right)}{(1-\pi_{case})} 
  \\ h'  E_1 X_i^f(D_i-p_i) \end{pmatrix}$ for $i=1, \ldots, n$ are independent and identically distributed with mean $\mathbf{0}$ and variance $\Sigma_f^*$, for some $\Sigma_f^*$ to be derived later. Hence, by multivariate Central limit Theorem, we get, \begin{equation}
 \frac{1}{n} \sum_{i=1}^n \begin{pmatrix}\frac{I(D_i=1)\left(I(X_i^{f'}\beta_f^* >u) - P(X^{f'}\beta_f^* >u) \right)}{\pi_{case}} 
  \\ g' E_1 X_i^f(D_i-p_i)  
  \\ \frac{I(D_i=0)\left(I(X_i^{f'}\beta_f^*\leq u)- (1-t)\right)}{(1-\pi_{case})} 
  \\ h'  E_1 X_i^f(D_i-p_i) \end{pmatrix}   \underset{\rightarrow}{d} \ \  \mathcal{N}(0, \Sigma_f^*)
\end{equation} 

The next step is to compute $\Sigma_f^*$. Thus, we need to compute the individual variances $Var(A_i)$ for $i=1,2,3,4$ and the covariances $Cov(A_i, A_j)$ for $i=1, \ldots 4, j=1, \ldots 4, i \neq j$.

\begin{align*}Var(A_1) &= Var\left(\frac{1}{n_1} \sum_{i:D_i=1} \left(I(X_i^{f'}\beta_f^* >u) - P(X^{f'}\beta_f^* >u)  \right) \right) = \frac{P(X^{f'}\beta_f^* >u)(1-P(X^{f'}\beta_f^* >u))}{n_1}\end{align*}

\begin{align*}&Var(A_2) = Var\left(\sum_{i=1}^n g'E_1X_i^f(D_i-p_i)  \right) = g' E_1' Var\left(\sum_{i=1}^nX_i^f(D_i-p_i) \right) E_1g\end{align*}

\begin{align*}
    &Var(A_3) = Var \left(\frac{{f}_{D_1,f}({{\beta}}_f^*,u)}{n_0 {f}_{D_0,f}({{\beta}}_f^*,u)} \sum_{i:D_i=0} \left(I(X_i^{f'}\beta_f^*\leq u)- (1-t)\right)\right)= \frac{{f}_{D_1,f}^{2}({{\beta}}_f^*,u)t(1-t)}{n_0 {f}_{D_0,f}^{2}({{\beta}}_f^*,u)} 
\end{align*}

\begin{align*}&Var(A_4) = Var\left(\sum_{i=1}^n h' E_1'X_i^f(D_i-p_i)  \right) \\ &= h'E_1' Var\left(\sum_{i=1}^nX_i^f(D_i-p_i) \right)E_1 h\end{align*}

We now look at the covariance terms one by one. First note, $Cov(A_i, A_j)=E(A_jA_j)$ for all $i \neq j$, as we have $E(A_i)=0$ for $i=1,2,3,4$.

\begin{align*}
    Cov(A_1, A_2) 
   &= E\left[\left(\frac{1}{n_1} \sum_{i:D_i=1} \left(I(X_i^{f'}\beta_f^* >u) - P(X^{f'}\beta_f^* >u)  \right) \right) \left(\sum_{j=1}^n g'E_1'X_j^f(D_j-p_j)  \right) \right] \\
   &= \frac{1}{n_1} \left[\sum_{i:D_i=1} \sum_{j=1}^n E\left(\left(I(X_i^{f'}\beta_f^* >u) - P(X^{f'}\beta_f^* >u)  \right) g'E_1'X_j^f(D_j-p_j)\right)\right] \\ &=\frac{1}{n_1} \left[\sum_{i:D_i=1}  E\left(\left(I(X_i^{f'}\beta_f^* >u) - P(X^{f'}\beta_f^* >u)  \right) g'E_1'X_i^f(D_i-p_i)\right)\right] \\ &\text{ (All $j\neq i$ terms vanish)}
\end{align*}

\begin{equation*}
    Cov(A_1, A_3) = 0 \text{ (Since $A_1$ and $A_3$ are purely dependent on cases and controls respectively)}.
\end{equation*}

\begin{align*}
     Cov(A_1, A_4)  &= E\left[\left(\frac{1}{n_1} \sum_{i:D_i=1} \left(I(X_i^{f'}\beta_f^* >u) - P(X^{f'}\beta_f^* >u)  \right) \right) \left(\sum_{j=1}^n h'E_1'X_j^f(D_j-p_j)  \right) \right] \\
   &= \frac{1}{n_1} \left[\sum_{i:D_i=1} \sum_{j=1}^n E\left(\left(I(X_i^{f'}\beta_f^* >u) - P(X^{f'}\beta_f^* >u)  \right) h'E_1'X_j^f(D_j-p_j)\right)\right] \\ &=\frac{1}{n_1} \left[\sum_{i:D_i=1}  E\left(\left(I(X_i^{f'}\beta_f^* >u) - P(X^{f'}\beta_f^* >u)  \right) h'E_1'X_i^f(D_i-p_i)\right)\right]
\end{align*}

\begin{align*}
    Cov(A_2, A_3) &= E\left[\left(\sum_{j=1}^n g'E_1'X_j^f(D_j-p_j)  \right) \left( \frac{{f}_{D_1,f}({{\beta}}_f^*,u)}{n_0 {f}_{D_0,f}({{\beta}}_f^*,u)} `\sum_{i:D_i=0} \left(I(X_i^{f'}\beta_f^*\leq u)- (1-t)\right) \right)  \right] \\
    &= \frac{{f}_{D_1,f}({{\beta}}_f^*,u)}{n_0 {f}_{D_0,f}({{\beta}}_f^*,u)}  \sum_{i:D_i=0} E \left(\left(I(X_i^{f'}\beta_f^*\leq u)- (1-t)\right)g'E_1'X_i^f(D_i-p_i) \right)
\end{align*}

\begin{align*}
    Cov(A_2, A_4) &= g'E_1' Var(\sum_{i=1}^n X_i^f(D_i-p_i))E_1 h
\end{align*}

\begin{align*}
    Cov(A_3, A_4) &= E \left[ \left( \frac{{f}_{D_1,f}({{\beta}}_f^*,u)}{n_0 {f}_{D_0,f}({{\beta}}_f^*,u)} \sum_{i:D_i=0} \left(I(X_i^{f'}\beta_f^*\leq u)- (1-t)\right) \right) \left(\sum_{j=1}^n h'E_1'X_j^f(D_j-p_j)  \right)\right] \\
    &= \frac{{f}_{D_1,f}({{\beta}}_f^*,u)}{n_0 {f}_{D_0,f}({{\beta}}_f^*,u)} \sum_{i:D_i=0} E \left(\left(I(X_i^{f'}\beta_f^*\leq u)- (1-t)\right)h'E_1'X_i^f(D_i-p_i)  \right)
\end{align*}

Thus, we get, \begin{eqnarray*}
    \begin{pmatrix}
        A_1\\ A_2 \\ A_3 \\ A_4 
    \end{pmatrix} \xrightarrow{d} \mathcal{N}\left(0, \frac{\Sigma_f^*}{n}\right)
\end{eqnarray*} 

Noting that, \begin{eqnarray*}
   \widehat{ROC}_f(t)-ROC_f(t)= A_1+A_2+A_3+A_4= \mathbf{1}' \begin{pmatrix}
        A_1\\ A_2 \\ A_3 \\ A_4 
    \end{pmatrix} 
\end{eqnarray*} we finally get,

\begin{equation*}
    \widehat{ROC}_f(t)-ROC_f(t) \stackrel{d}{\rightarrow} \mathcal{N}(0, \Sigma_f)
\end{equation*} where $\Sigma_f= \frac{1}{n}\mathbf{1}'\Sigma_f^*\mathbf{1}$

\section{Proof of Theorem 2:}
In Appendix A, we have derived the expression for $\Sigma_f$. In a similar way, we can also derive the expression for $\Sigma_r$. Next, we are interested in incremental performance.  Hence, we now look at the quantity $[\widehat{ROC}_f(t)-\widehat{ROC}_r(t)]-[ROC_f(t)-ROC_r(t)]$. In order to prove that this quantity is asymptotically normal, we first need to show that $\begin{pmatrix}
         \widehat{ROC}_f(t)-ROC_f(t) \\   \widehat{ROC}_r(t)-ROC_r(t) 
    \end{pmatrix} $ is bivariate normal.

In order to argue $
    \begin{pmatrix}
         \widehat{ROC}_f(t)-ROC_f(t) \\   \widehat{ROC}_r(t)-ROC_r(t) 
    \end{pmatrix} 
$ is bivariate normal, we note that \begin{align*}
     &\begin{pmatrix}
         \widehat{ROC}_f(t)-ROC_f(t) \\   \widehat{ROC}_r(t)-ROC_r(t) 
    \end{pmatrix} \\ &= \frac{1}{n} \sum_{i=1}^n \begin{pmatrix}
        \frac{I(D_i=1)\left(I(X_i^{f'}\beta_f^* >u) - P(X^{f'}\beta_f^* >u) \right)}{\pi_{case}}   
  + \frac{I(D_i=0)\left(I(X_i^{f'}\beta_f^*\leq u)- (1-t)\right)}{(1-\pi_{case})} 
  + (g+h)'  E_1 X_i^f(D_i-p_i) \\    \frac{I(D_i=1)\left(I(X_i^{r'}\beta_r >v) - P(X^{r'}\beta_r >v) \right)}{\pi_{case}}   
  + \frac{I(D_i=0)\left(I(X_i^{r'}\beta_r\leq v)- (1-t)\right)}{(1-\pi_{case})} 
  + (g_1+h_1)'  E_2 X_i^r(D_i-p_i)
    \end{pmatrix}
\end{align*} is the sum of independent and identically distributed random vectors, where $v, E_2, g_1, h_1$ are defined analogously to $u, E_1, g, h$ for the restricted model. Hence, by the Multivariate Central Limit Theorem, we have \begin{equation*}
    \begin{pmatrix}
         \widehat{ROC}_f(t)-ROC_f(t) \\   \widehat{ROC}_r(t)-ROC_r(t) 
    \end{pmatrix} \stackrel{d}{\rightarrow} \mathcal{N}(0, \Sigma^*)
\end{equation*} where $\Sigma^*= \begin{pmatrix}
    \Sigma_f & \Sigma_{fr} \\ \Sigma_{fr} & \Sigma_r
\end{pmatrix}$ The expressions of $\Sigma_f$ and $\Sigma_r$ have been defined before. For $\Sigma_{fr}$, we note that \begin{eqnarray*}
    \Sigma_{fr}= &\frac{1}{n}\sum_{i=1}^n E \bigl(\left( \frac{I(D_i=1)\left(I(X_i^{f'}\beta_f^* >u) - P(X^{f'}\beta_f^* >u) \right)}{\pi_{case}}   
  + \frac{I(D_i=0)\left(I(X_i^{f'}\beta_f^*\leq u)- (1-t)\right)}{(1-\pi_{case})} 
  + (g+h)'  E_1 X_i^f(D_i-p_i) \right)\\ &\left( \frac{I(D_i=1)\left(I(X_i^{r'}\beta_r >v) - P(X^{r'}\beta_r >v) \right)}{\pi_{case}}   
  + \frac{I(D_i=0)\left(I(X_i^{r'}\beta_r\leq v)- (1-t)\right)}{(1-\pi_{case})} 
  + (g_1+h_1)'  E_2 X_i^r(D_i-p_i) \right)\bigr)\end{eqnarray*}

\noindent Hence, we finally get \begin{equation*}
   [\widehat{ROC}_f(t)-\widehat{ROC}_r(t)]-[ROC_f(t)-ROC_r(t)] \stackrel{d}{\rightarrow} \mathcal{N}(0,  \Sigma)
\end{equation*} where $\Sigma = \Sigma_f +\Sigma_r -2 \Sigma_{fr}$
\clearpage
\section{Results of Simulation Studies}

\subsection{Probabilities of rejection under a correctly specified model for $t=0.1$}
\begin{table}[!htbp]
  \caption{Probabilities of rejection in Stage 1, Stage 2, combined probabilities of rejection, and the probabilities of continuing to Stage 2 for the correctly specified model and for two boundary choices: O'Brien Fleming \& Pocock, with 200 cases, 200 controls; and 200 cases, 400 controls for $t=0.1, \lambda= 1/2 \ \& \   \delta_0=0.141$.  Here, we stop for both futility and efficacy.}
    \centering
    \resizebox{\textwidth}{!}{
        \begin{tabular}{cccccccc}
            \hline
            Boundary & Case & $\mu_{f}$ & $\delta$ & $S1: P(Reject|H)$ & $S1: P(Continue|H)$ & $S2: P(Reject|H)$ & $P(Reject|H)$ \\ \midrule
          \multicolumn{8}{c}{200 cases and 200 controls} \\ \hline
            \multirow{3}{*}{O'Brien Fleming} 
            & Null & $(1,1.1)$ & 0.141 & 0.015 (0.002) & 0.480 (0.007) & 0.042 (0.003) & 0.056 (0.003) \\ 
            & \multirow{2}{*}{Alternate} 
            & (1,1.5) & 0.259 & 0.171 (0.005) & 0.721 (0.006) & 0.415 (0.007) & 0.586 (0.007) \\ 
            &  & (0.8, 2) & 0.461 & 0.877 (0.005) & 0.123 (0.005) & 0.123 (0.005) & 0.999 (0.000) \\ \hline
            
            \multirow{3}{*}{Pocock} 
            & Null & $(1,1.1)$ & 0.141 & 0.040 (0.003) & 0.173 (0.005) & 0.014 (0.002) & 0.054 (0.003) \\ 
            & \multirow{2}{*}{Alternate} 
            & (1,1.5) & 0.259 & 0.334 (0.007) & 0.395 (0.007) & 0.204 (0.006) & 0.537 (0.007) \\ 
            &  & (0.8,2) & 0.461 & 0.947 (0.003) & 0.048 (0.003) & 0.048 (0.003) & 0.996 (0.001) \\ \hline
            \multicolumn{8}{c}{200 cases and 400 controls} \\ \hline  
            \multirow{3}{*}{O'Brien Fleming} 
            & Null & $(1,1.1)$ & 0.141 & 0.013 (0.002) & 0.484 (0.007) & 0.038 (0.003) & 0.051 (0.003) \\ 
            & \multirow{2}{*}{Alternate} 
            & (1,1.5) & 0.259 & 0.243 (0.006) & 0.695 (0.007) & 0.489 (0.007) & 0.733 (0.006) \\ 
            &  & (0.8, 2) & 0.461 & 0.967 (0.003) & 0.033 (0.003) & 0.032 (0.003) & 1.000 (0.000) \\ \hline
            
            \multirow{3}{*}{Pocock} 
            & Null & $(1,1.1)$ & 0.141 & 0.036 (0.003) & 0.191 (0.006) & 0.021 (0.002) & 0.057 (0.003) \\ 
            & \multirow{2}{*}{Alternate} 
            & (1,1.5) & 0.259 & 0.419 (0.007) & 0.381 (0.007) & 0.246 (0.006) & 0.665 (0.007) \\ 
            &  & (0.8, 2) & 0.461 & 0.991 (0.001) & 0.009 (0.001) & 0.009 (0.001) & 1.000 (0.000) \\ \hline
        \end{tabular}
    }
    \label{tab:prob_0.1}
\end{table}

\begin{table}[!htbp]
\caption{Probabilities of rejection in Stage 1, Stage 2, combined probabilities of rejection, and the probabilities of continuing to Stage 2 for the correctly specified model and for two boundary choices: O'Brien Fleming \& Pocock, with 200 cases, 200 controls; and 200 cases, 400 controls for $t=0.1, \lambda=1/2 \ \& \ \delta_0=0.141$. Here, we stop only for futility. }
    \centering
    \resizebox{\textwidth}{!}{
        \begin{tabular}{@{}cccccccc@{}}
            \hline
            Boundary & Case & $\mu_{f}$ & $\delta$ & $S1: P(Reject|H)$ & $S1: P(Continue|H)$ & $S2: P(Reject|H)$ & $P(Reject|H)$ \\ \midrule
          \multicolumn{8}{c}{200 cases and 200 controls} \\ \hline
            \multirow{3}{*}{O'Brien Fleming} & Null & $(1,1.1)$ & 0.141 & 0(0) & 0.526 (0.007) & 0.059 (0.003) & 0.059 (0.003) \\ 
            & \multirow{2}{*}{Alternate} & (1, 1.5) & 0.259 & 0(0) & 0.917 (0.004) & 0.590 (0.007) & 0.590 (0.007) \\ 
            & & (0.8, 2) & 0.461 & 0(0) & 1.000 (0.000) & 1.000 (0.000) & 1.000 (0.000) \\ \hline \multirow{3}{*}{Pocock} 
        & Null & $(1,1.1)$ & 0.141 & 0(0) & 0.291 (0.006) & 0.056 (0.003) & 0.056 (0.003) \\ 
        & \multirow{2}{*}{Alternate} 
        & (1, 1.5) & 0.259 & 0(0) & 0.771 (0.006) & 0.558 (0.007) & 0.558 (0.007) \\ 
        &  & (0.8, 2) & 0.461 & 0(0) & 0.999 (0.000) & 0.998 (0.001) & 0.998 (0.001) \\ \hline
        \multicolumn{8}{c}{200 cases and 400 controls} \\ \hline
        \multirow{3}{*}{O'Brien Fleming} 
        & Null & $(1,1.1)$ & 0.141 & 0(0) & 0.532 (0.007) & 0.057 (0.003) & 0.057 (0.003) \\ 
        & \multirow{2}{*}{Alternate} 
        & (1, 1.5) & 0.259 & 0(0) & 0.957 (0.003) & 0.736 (0.006) & 0.736 (0.006) \\ 
        &  & (0.8, 2) & 0.461 & 0(0) & 1.000 (0.000) & 1.000 (0.000) & 1.000 (0.000) \\ \hline
        
        \multirow{3}{*}{Pocock} 
        & Null & $(1,1.1)$ & 0.141 & 0(0) & 0.300 (0.006) & 0.049 (0.003) & 0.049 (0.003) \\ 
        & \multirow{2}{*}{Alternate} 
        & (1, 1.5) & 0.259 & 0(0) & 0.842 (0.005) & 0.702 (0.006) & 0.702 (0.006) \\ 
        &  & (0.8, 2) & 0.461 & 0(0) & 1.000 (0.000) & 1.000 (0.000) & 1.000 (0.000) \\ \hline
    \end{tabular}
}
\label{tab:prob_0.1_fut}
\end{table}

\clearpage
\subsection{Comparison of analytical and simulation results for group rotation for $t=0.1$ under a correctly specified model}

\begin{figure}[b]
    \centering
    \includegraphics[width=6cm , trim=0 0 4cm 0, clip]{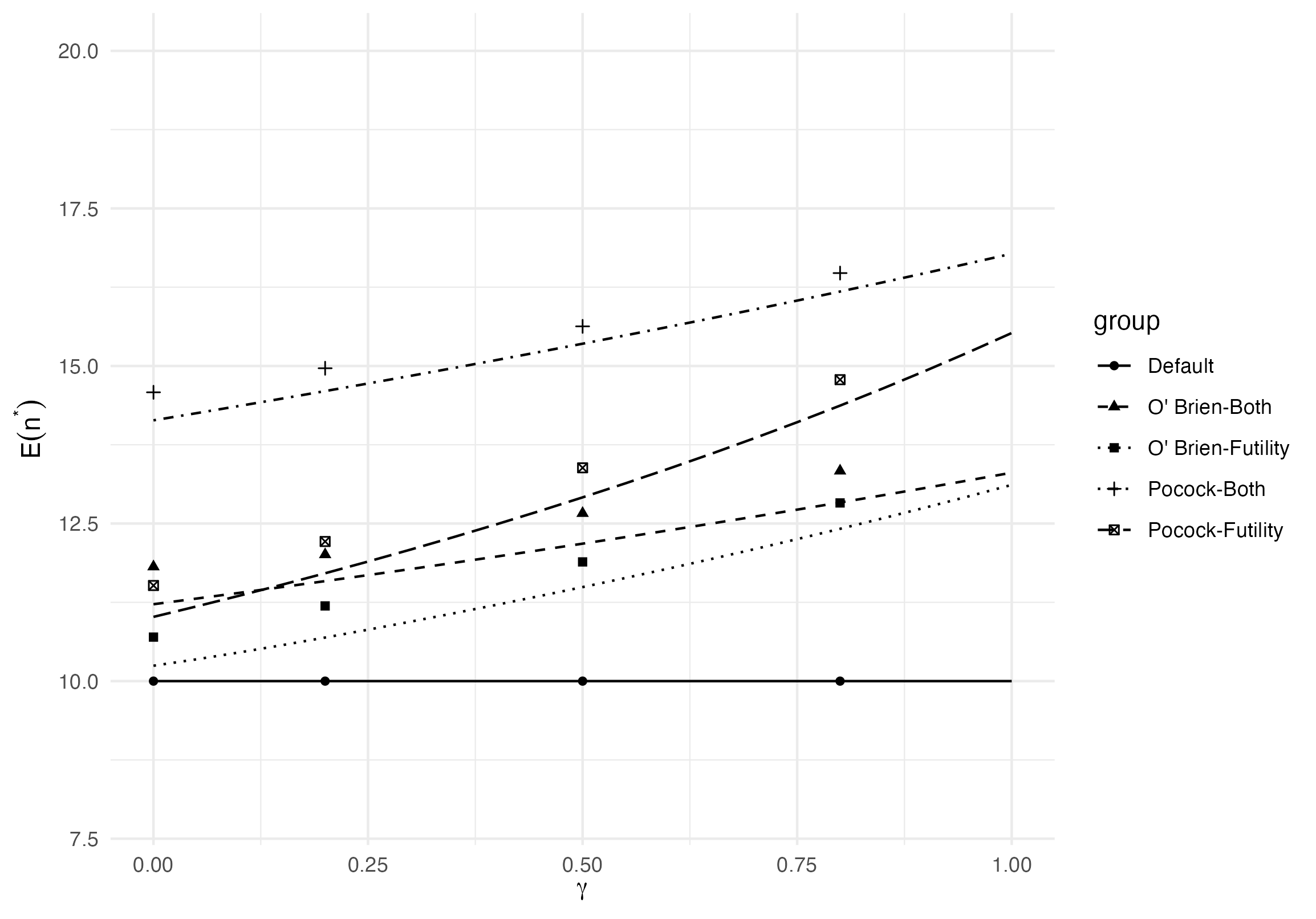}
\includegraphics[width=7.5cm]{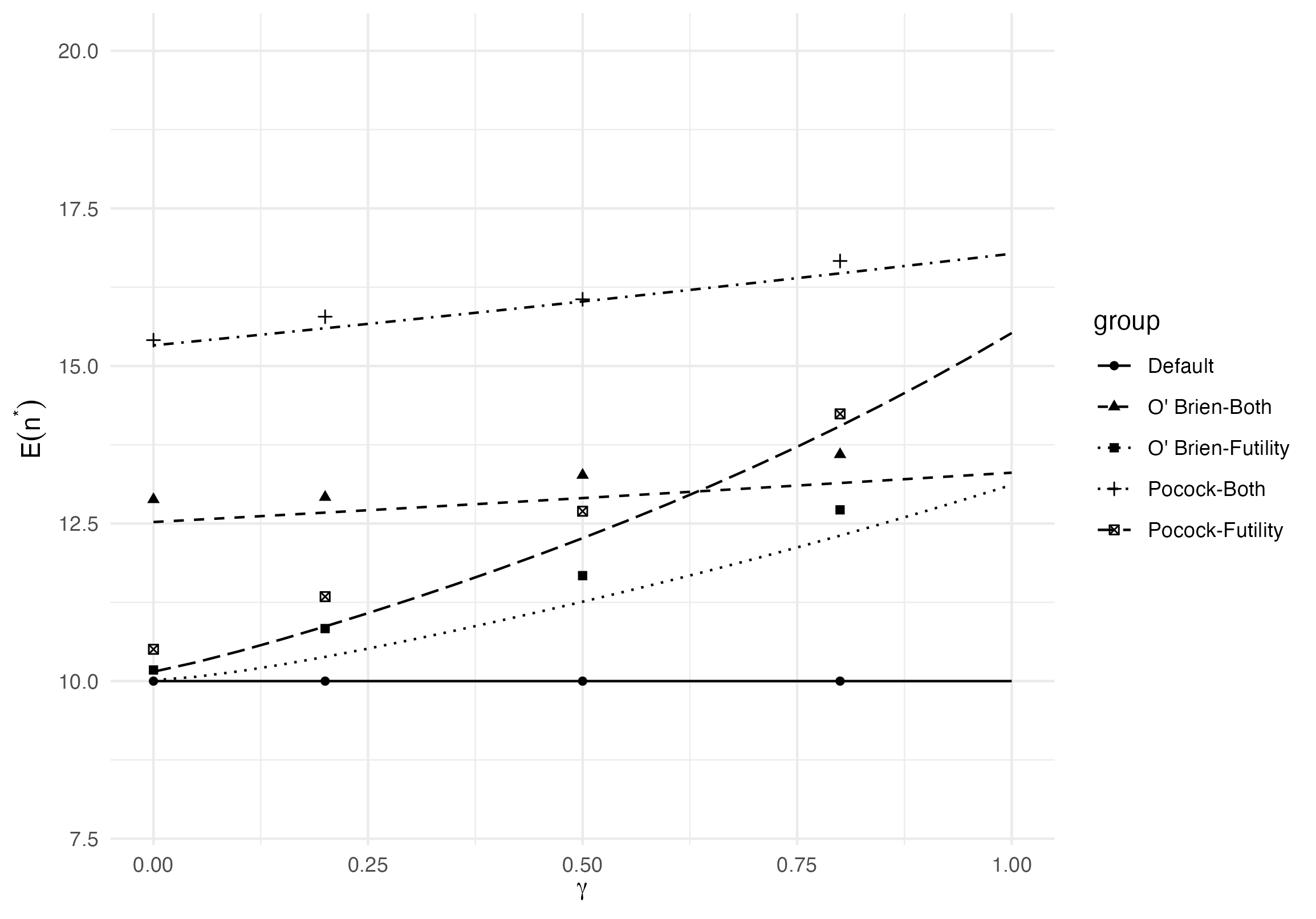}
 \includegraphics[width=6cm , trim=0 0 4cm 0, clip]{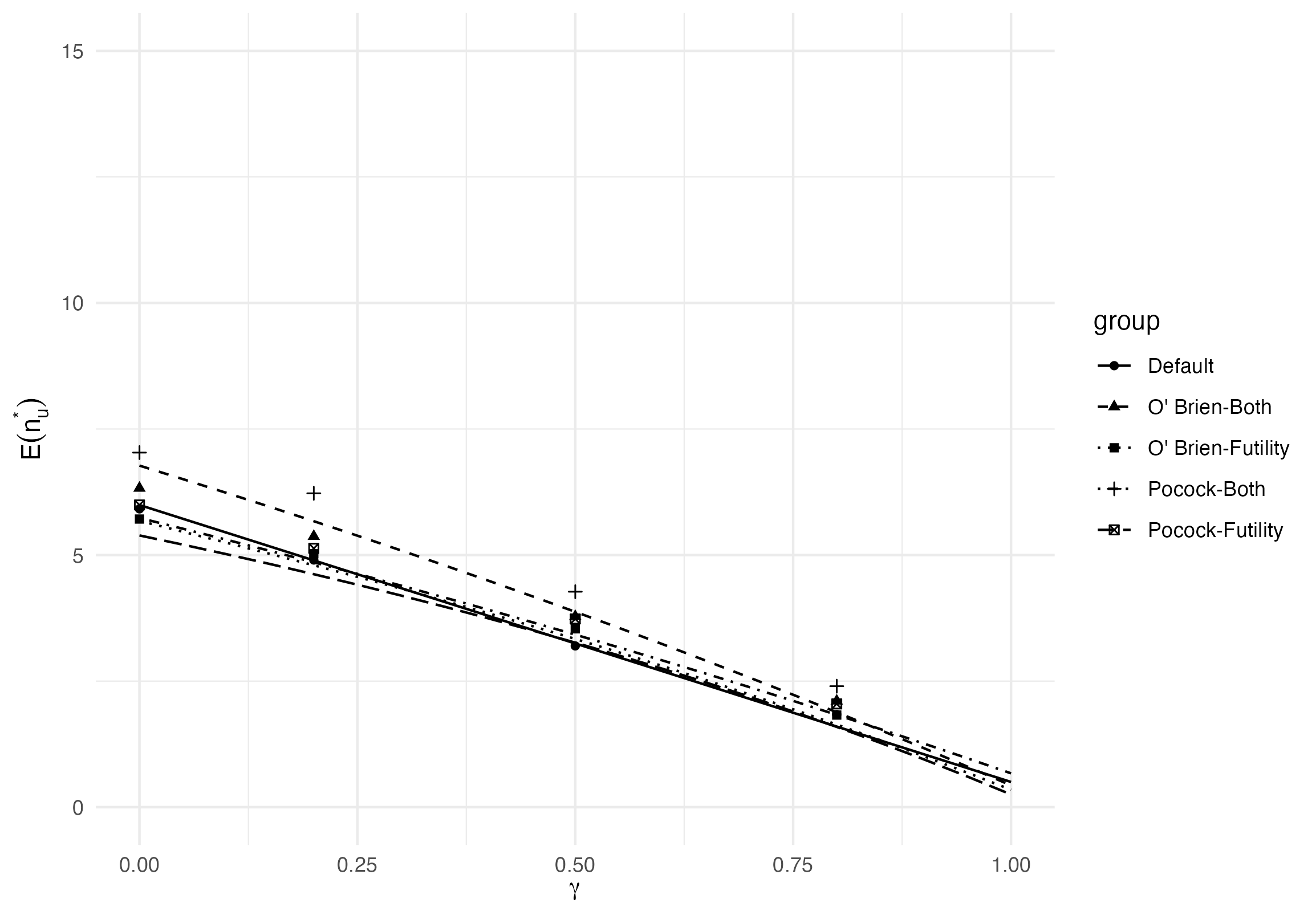}
\includegraphics[width=7.5cm]{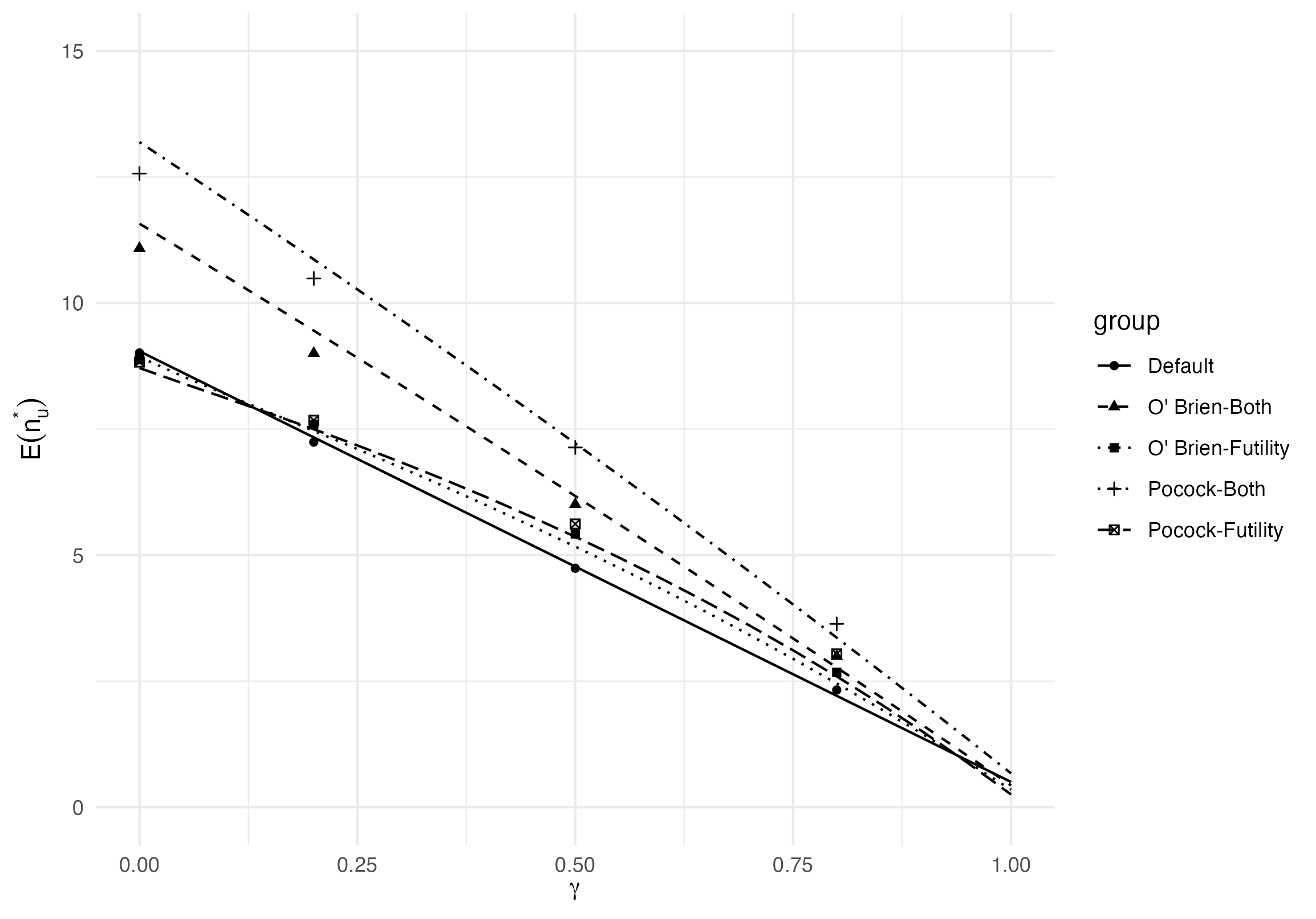}
 \includegraphics[width=6cm , trim=0 0 4cm 0, clip]{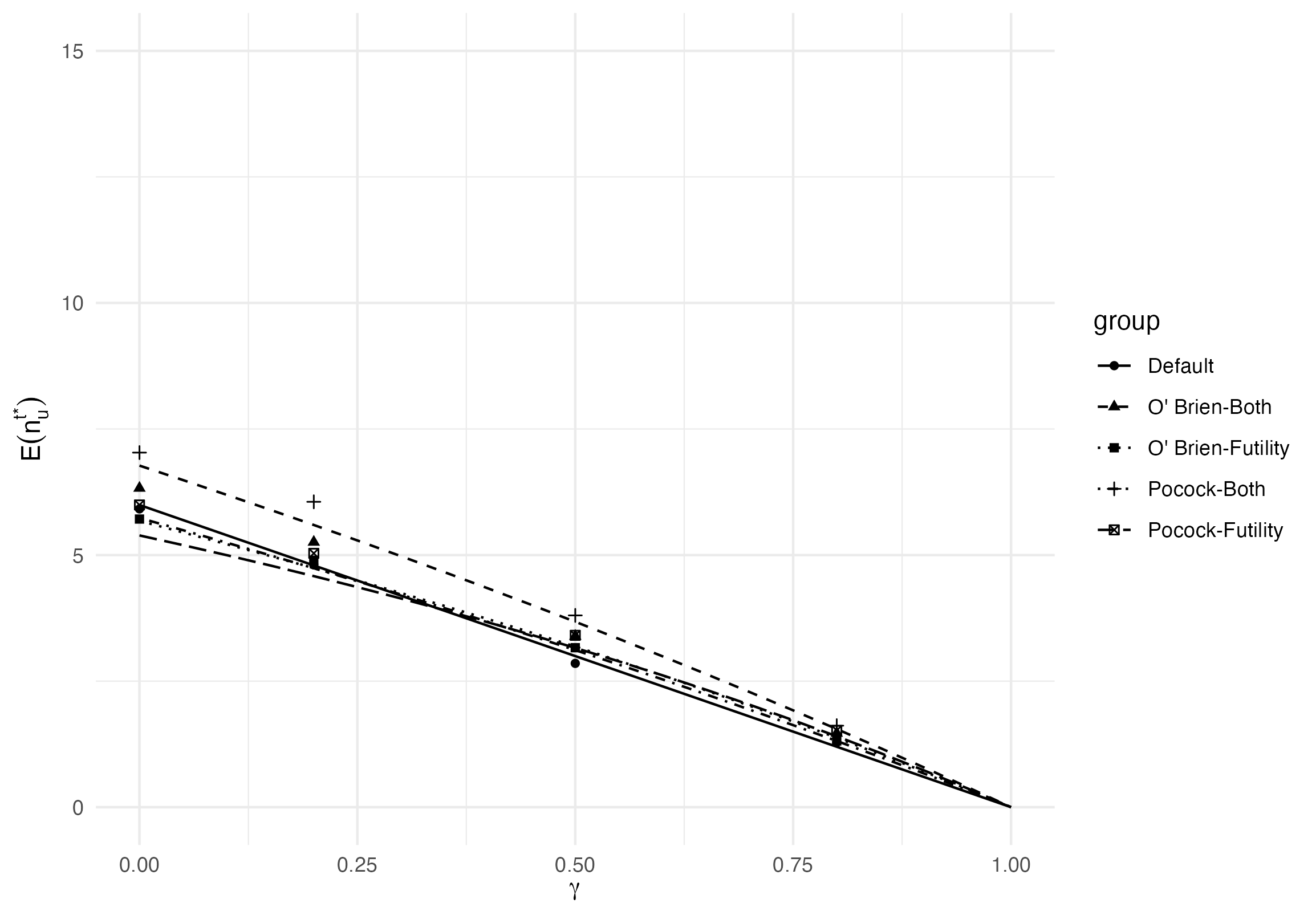} 
     \includegraphics[width=7.5cm]{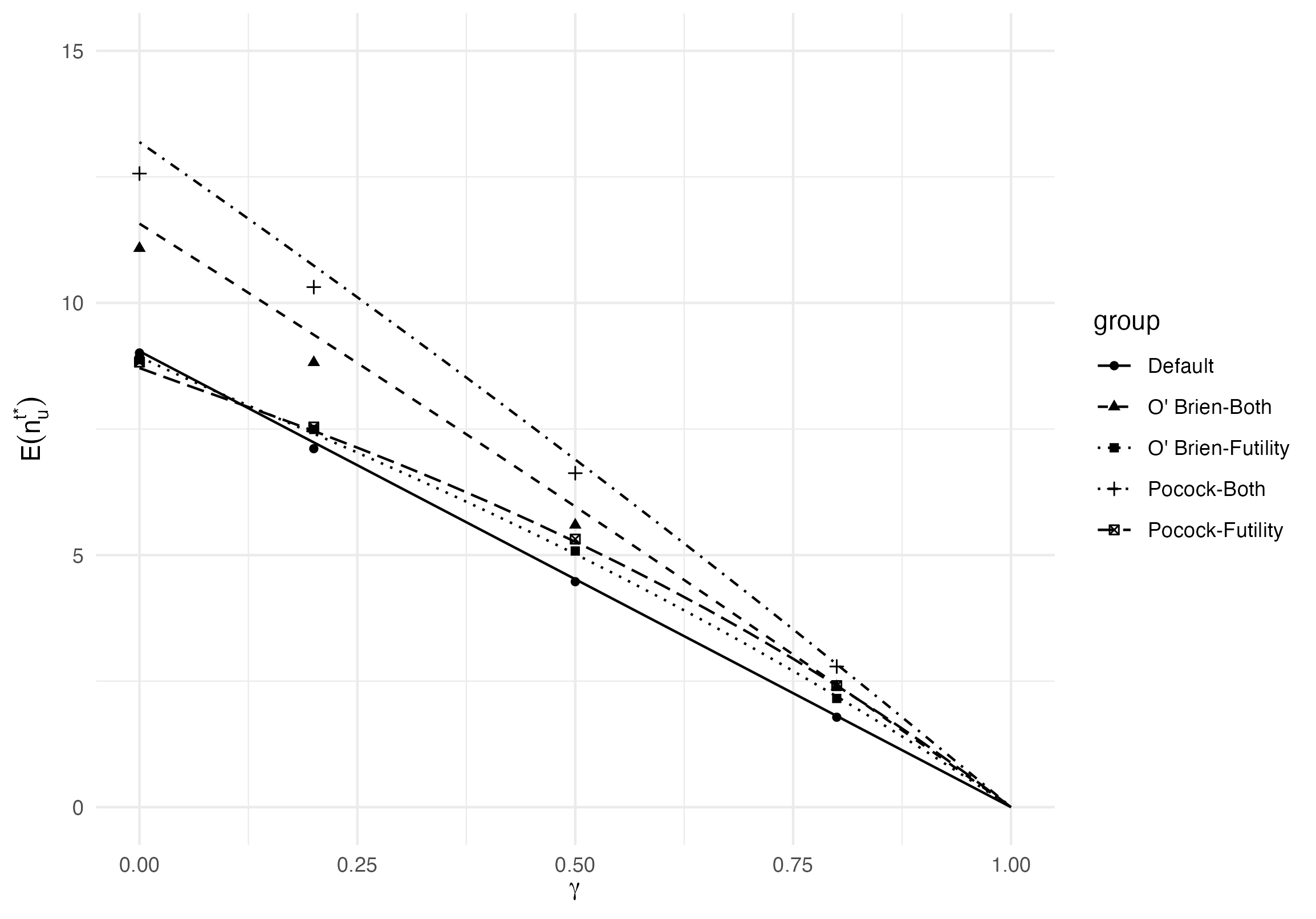}
    \caption{Comparison of the values of $E(n^*), E(n_u^*), E(n_u^{t*})$ computed based on the analytical formula and simulation for $t=0.1$ for the correctly specified model. We consider two boundary criteria: O' Brien Fleming and Pocock and two stopping criteria: Stopping for both Futility and Efficacy and Stopping only for Futility. Results across combinations of these boundary and stopping criteria are compared with results from a non-sequential procedure, denoted as `Default'. The lines denote the values obtained from the analytical formula, while the dots in several shapes denote the simulation results at $\gamma=0, 0.2, 0.5, 0.8$. The figures to the left correspond to 200 cases and 200 controls, while the figures to the right correspond to 500 cases and 500 controls. }
    \label{fig:rotation_1_eq}
\end{figure}

\begin{table}[!htbp]
 \caption{Comparison of the values of $E(n^*), E(n_u^*), E(n_u^{t*})$ computed based on the analytical formula and simulation for $t=0.1$ based on 500 cases and 500 controls for the correctly specified model. We consider two boundary criteria: O' Brien Fleming and Pocock and two stopping criteria: Stopping for both Futility and Efficacy and Stopping only for Futility. Results across combinations of these boundary and stopping criteria are compared with results from a non-sequential procedure, denoted as `Default'. We consider four different values of the mixing parameter $\gamma=0, 0.2, 0.5, 0.8$. The values in brackets denote the corresponding standard errors based on 2000 simulations.}
    \centering
   \resizebox{\textwidth}{!}{  \begin{tabular}{ccccccccc}
    \hline
       Choice of $\gamma$ & Stop & & \multicolumn{2}{c}{$E(n^*)$ } & \multicolumn{2}{c}{$E(n_u^*)$} & \multicolumn{2}{c}{$E(n_u^{t*})$ } \\ \hline
    &  &  & True & Estimate  & True & Estimate  & True & Estimate \\ \midrule
\multirow{5}{*}{$\gamma=0$ } & \multirow{2}{*}{Both } & O' Brien Fleming & 12.524 & 12.882 (0.051) & 11.572 & 11.084 (0.076) & 11.572 & 11.084 (0.076) \\ 
& & Pocock & 15.327 & 15.410 (0.060) & 13.190 & 12.566 (0.088) & 13.190 & 12.566 (0.088) \\
& \multirow{2}{*}{Futility } & O' Brien Fleming & 10.013 & 10.176 (0.017) & 8.929 & 8.874 (0.044) & 8.929 & 8.874 (0.044) \\
& & Pocock & 10.149 & 10.505 (0.025) & 8.709 & 8.828 (0.045) & 8.709 & 8.828 (0.045) \\
& Default & & 10.000 & 10.000 (0.000) & 9.042 & 9.008 (0.043) & 9.042 & 9.008 (0.043) \\  \hline

\multirow{5}{*}{$\gamma=0.2$ } & \multirow{2}{*}{Both } & O' Brien Fleming & 12.674 & 12.918 (0.047) & 9.451 & 9.002 (0.082) & 9.368 & 8.820 (0.083) \\ 
& & Pocock & 15.598 & 15.782 (0.060) & 10.864 & 10.486 (0.093) & 10.739 & 10.314 (0.094) \\
& \multirow{2}{*}{Futility } & O' Brien Fleming & 10.384 & 10.834 (0.026) & 7.463 & 7.593 (0.060) & 7.408 & 7.491 (0.061) \\
& & Pocock & 10.867 & 11.341 (0.035) & 7.496 & 7.673 (0.056) & 7.461 & 7.539 (0.058)\\
& Default & & 10.000 & 10.000 (0.000) & 7.334 & 7.240 (0.062) & 7.234 & 7.110 (0.064) \\  \hline

\multirow{5}{*}{$\gamma=0.5$ } & \multirow{2}{*}{Both } & O' Brien Fleming & 12.904 & 13.268 (0.056) & 6.171 & 6.006 (0.079) & 5.962 & 5.596 (0.079) \\ 
& & Pocock & 12.904 & 16.058 (0.062) & 6.171 & 7.134 (0.092) & 5.962 & 6.624 (0.088) \\
& \multirow{2}{*}{Futility } & O' Brien Fleming & 11.260 & 11.673 (0.039) & 5.169 & 5.419 (0.067) & 5.020 & 5.080 (0.069) \\
& & Pocock & 11.260 & 12.695 (0.049) & 5.169 & 5.617 (0.068) & 5.020 & 5.317 (0.069) \\
& Default & & 10.000 & 10.000 (0.000) & 4.771 & 4.738 (0.073) & 4.521 & 4.472 (0.072)  \\  \hline

\multirow{5}{*}{$\gamma=0.8$ } & \multirow{2}{*}{Both } & O' Brien Fleming & 13.143 & 13.596 (0.055) & 2.770 & 2.992 (0.069) & 2.429 & 2.392 (0.063)  \\ 
& & Pocock & 16.470 & 16.666 (0.063) & 3.362 & 3.636 (0.081) & 2.835 & 2.790 (0.070) \\
& \multirow{2}{*}{Futility } & O' Brien Fleming & 12.308 & 12.717 (0.050) & 2.455 & 2.675 (0.060) & 2.195 & 2.156 (0.055) \\
& & Pocock & 14.047 & 14.240 (0.057) & 2.593 & 3.040 (0.063) & 2.411 & 2.411 (0.058) \\
& Default & & 10.000 & 10.000 (0.000) & 2.208 & 2.324 (0.058) & 1.808 & 1.784 (0.054) \\  \hline
    \end{tabular}}
    \label{tab:sim_rotation_500_eq}
\end{table}

\begin{table}[!htbp]
 \caption{Comparison of the values of $E(n^*), E(n_u^*), E(n_u^{t*})$ computed based on the analytical formula and simulation for $t=0.1$ based on 200 cases and 200 controls for the correctly specified model. We consider two boundary criteria: O' Brien Fleming and Pocock and two stopping criteria: Stopping for both Futility and Efficacy and Stopping only for Futility. Results across combinations of these boundary and stopping criteria are compared with results from a non-sequential procedure, denoted as `Default'. We consider four different values of the mixing parameter $\gamma=0, 0.2, 0.5, 0.8$. The values in brackets denote the corresponding standard errors based on 2000 simulations.}
    \centering
   \resizebox{\textwidth}{!}{  \begin{tabular}{ccccccccc}
    \hline
       Choice of $\gamma$ & Stop & & \multicolumn{2}{c}{$E(n^*)$ } & \multicolumn{2}{c}{$E(n_u^*)$} & \multicolumn{2}{c}{$E(n_u^{t*})$ } \\ \hline
    &  &  & True & Estimate  & True & Estimate  & True & Estimate \\ \midrule
\multirow{5}{*}{$\gamma=0$ } & \multirow{2}{*}{Both } & O' Brien Fleming &11.220 & 11.815 (0.040) & 6.774 & 6.327 (0.083) & 6.774 & 6.327 (0.083) \\ 
& & Pocock & 14.137 & 14.582 (0.058) & 5.736 & 7.032 (0.087) & 5.736 & 7.032 (0.087) \\
& \multirow{2}{*}{Futility } & O' Brien Fleming &10.244 & 10.698 (0.027) & 5.679 & 5.714 (0.067) & 5.679 & 5.714 (0.067)  \\
& & Pocock & 11.019 & 11.517 (0.038) & 5.390 & 5.994 (0.070) & 5.390 & 5.994 (0.070)\\
& Default & & 10.000 & 10.000 (0.000) & 5.993 & 5.916 (0.068) & 5.993 & 5.916 (0.068) \\  \hline

\multirow{5}{*}{$\gamma=0.2$ } & \multirow{2}{*}{Both } & O' Brien Fleming &11.587 & 12.008 (0.047) & 5.672 & 5.371 (0.076) & 5.597 & 5.257 (0.076) \\ 
& & Pocock & 14.602 & 14.964 (0.057) & 4.857 & 6.225 (0.089) & 4.740 & 6.056 (0.089) \\
& \multirow{2}{*}{Futility } & O' Brien Fleming & 10.691 & 11.193 (0.033) & 4.798 & 5.026 (0.072) & 4.741 & 4.897 (0.072) \\
& & Pocock & 11.711 & 12.215 (0.047) & 4.620 & 5.133 (0.069) & 4.582 & 5.034 (0.070) \\
& Default & & 10.000 & 10.000 (0.000) & 4.895 & 4.902 (0.074) & 4.795 & 4.806 (0.074) \\  \hline

\multirow{5}{*}{$\gamma=0.5$ } & \multirow{2}{*}{Both } & O' Brien Fleming & 12.181 & 12.663 (0.049) & 3.875 & 3.783 (0.073) & 3.677 & 3.392 (0.073) \\ 
& & Pocock & 12.181 & 15.629 (0.067) & 3.875 & 4.273 (0.076) & 3.677 & 3.803 (0.075) \\
& \multirow{2}{*}{Futility } & O' Brien Fleming & 11.492 & 11.891 (0.043) & 3.337 & 3.537 (0.065) & 3.185 & 3.163 (0.061) \\
& & Pocock & 11.492 & 13.384 (0.055) & 3.337 & 3.732 (0.069) & 3.185 & 3.408 (0.066) \\
& Default & & 10.000 & 10.000 (0.000) & 3.247 & 3.198 (0.065) & 2.997 & 2.852 (0.063) \\  \hline

\multirow{5}{*}{$\gamma=0.8$ } & \multirow{2}{*}{Both } & O' Brien Fleming & 12.834 & 13.335 (0.055) & 1.883 & 2.106 (0.058) & 1.550 & 1.468 (0.049) \\ 
& & Pocock & 16.181 & 16.474 (0.065) & 1.831 & 2.400 (0.065) & 1.313 & 1.616 (0.056) \\
& \multirow{2}{*}{Futility } & O' Brien Fleming & 12.416 & 12.827 (0.049) & 1.639 & 1.827 (0.055) & 1.377 & 1.314 (0.049) \\
& & Pocock & 14.371 & 14.783 (0.062) & 1.592 & 2.048 (0.056) & 1.406 & 1.521 (0.052) \\
& Default & & 10.000 & 10.000 (0.000) & 1.599 & 1.838 (0.054) & 1.199 & 1.284 (0.048) \\  \hline
    \end{tabular}}
    \label{tab:sim_rotation_200_eq}
\end{table}
\clearpage

\subsection{Comparison of analytical and simulation results for group rotation for $t=0.1$ under a misspecified model with 200 cases and 200 controls}
\begin{table}[!htbp]
 \caption{Comparison of the values of $E(n^*), E(n_u^*), E(n_u^{t*})$ computed based on the analytical formula and simulation for $t=0.1$ based on 200 cases and 200 controls for the misspecified model. We consider two boundary criteria: O' Brien Fleming and Pocock and two stopping criteria: Stopping for both Futility and Efficacy and Stopping only for Futility. Results across combinations of these boundary and stopping criteria are compared with results from a non-sequential procedure, denoted as `Default'. We consider four different values of the mixing parameter $\gamma=0, 0.2, 0.5, 0.8$. The values in brackets denote the corresponding standard errors based on 2000 simulations.}
    \centering
   \resizebox{\textwidth}{!}{  \begin{tabular}{ccccccccc}
    \hline
       Choice of $\gamma$ & Stop & & \multicolumn{2}{c}{$E(n^*)$ } & \multicolumn{2}{c}{$E(n_u^*)$} & \multicolumn{2}{c}{$E(n_u^{t*})$ } \\ \hline
    &  &  & True & Estimate  & True & Estimate  & True & Estimate \\ \midrule
\multirow{5}{*}{$\gamma=0$ } & \multirow{2}{*}{Both } & O' Brien Fleming & 11.185 & 11.743 (0.041) & 6.313 & 6.018 (0.080) & 6.313 & 6.018 (0.080)  \\ 
& & Pocock & 14.127 & 14.471 (0.060) & 4.871 & 6.867 (0.094) & 4.871 & 6.867 (0.094)\\
& \multirow{2}{*}{Futility } & O' Brien Fleming & 10.367 & 10.781 (0.030) & 4.881 & 5.586 (0.066) & 4.881 & 5.586 (0.066)\\
& & Pocock & 11.318 & 11.694 (0.039) & 4.583 & 5.801 (0.066) & 4.583 & 5.801 (0.066)  \\
& Default & & 10.000 & 10.000 (0.000) & 5.634 & 5.538 (0.071) & 5.634 & 5.538 (0.071) \\  \hline

\multirow{5}{*}{$\gamma=0.2$ } & \multirow{2}{*}{Both } & O' Brien Fleming & 11.557 & 12.061 (0.042) & 5.293 & 5.004 (0.076) & 5.218 & 4.871 (0.076) \\ 
& & Pocock & 14.593 & 14.952 (0.062) & 4.142 & 5.642 (0.086) & 4.025 & 5.475 (0.084) \\
& \multirow{2}{*}{Futility } & O' Brien Fleming & 10.811 & 11.241 (0.034) & 4.129 & 4.688 (0.070) & 4.072 & 4.551 (0.071)  \\
& & Pocock & 11.978 & 12.352 (0.048) & 3.919 & 4.841 (0.067) & 3.880 & 4.730 (0.068)  \\
& Default & & 10.000 & 10.000 (0.000) & 4.607 & 4.488 (0.070) & 4.507 & 4.386 (0.070) \\  \hline

\multirow{5}{*}{$\gamma=0.5$ } & \multirow{2}{*}{Both } & O' Brien Fleming & 12.161 & 12.642 (0.050) & 3.629 & 3.537 (0.071) & 3.432 & 3.174 (0.070)  \\ 
& & Pocock & 12.161 & 15.705 (0.062) & 3.629 & 4.008 (0.078) & 3.432 & 3.556 (0.076)  \\
& \multirow{2}{*}{Futility } & O' Brien Fleming & 11.578 & 12.018 (0.043) & 2.879 & 3.368 (0.066) & 2.726 & 3.062 (0.063) \\
& & Pocock & 11.578 & 13.441 (0.052) & 2.879 & 3.688 (0.065) & 2.726 & 3.336 (0.066)\\
& Default & & 10.000 & 10.000 (0.000) & 3.067 & 3.010 (0.066) & 2.817 & 2.690 (0.063)\\  \hline

\multirow{5}{*}{$\gamma=0.8$ } & \multirow{2}{*}{Both } & O' Brien Fleming & 12.825 & 13.297 (0.055) & 1.780 & 2.040 (0.057) & 1.448 & 1.457 (0.049)  \\ 
& & Pocock & 16.179 & 16.473 (0.065) & 1.633 & 2.259 (0.063) & 1.116 & 1.503 (0.052) \\
& \multirow{2}{*}{Futility } & O' Brien Fleming & 12.456 & 12.873 (0.050) & 1.436 & 1.901 (0.057) & 1.173 & 1.352 (0.049)  \\
& & Pocock & 14.467 & 14.807 (0.062) & 1.359 & 2.006 (0.057) & 1.172 & 1.441 (0.050)\\
& Default & & 10.000 & 10.000 (0.000) & 1.527 & 1.552 (0.051) & 1.127 & 1.084 (0.046)\\  \hline
    \end{tabular}}
    \label{tab:sim_rotation_200_uneq}
\end{table}
\clearpage
\subsection{Comparison of analytical and simulation results for group rotation for $t=0.2$ under a misspecified model}

\begin{figure}[!htbp]
    \centering
    \includegraphics[width=6cm , trim=0 0 4cm 0, clip]{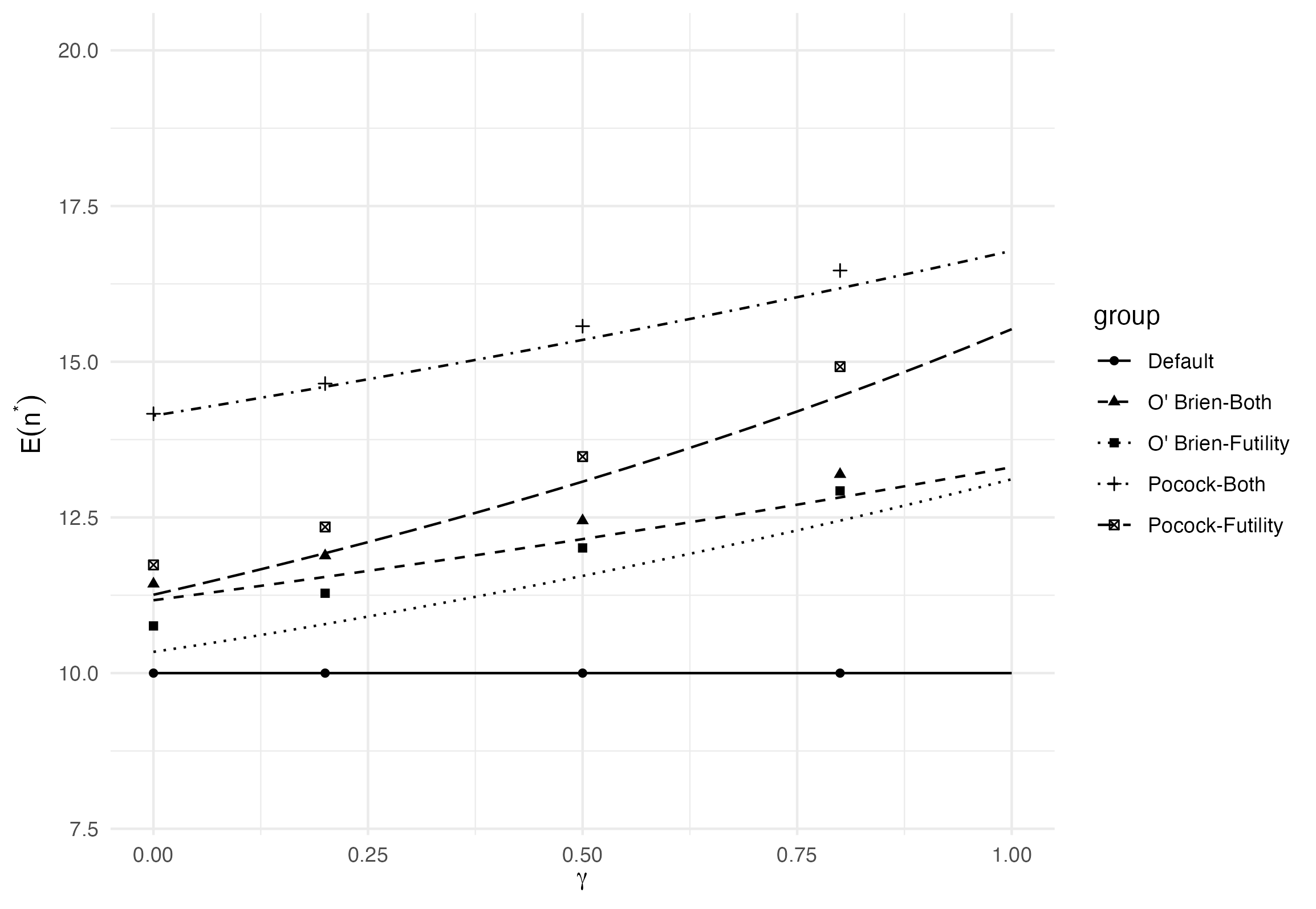}
\includegraphics[width=7.5cm]{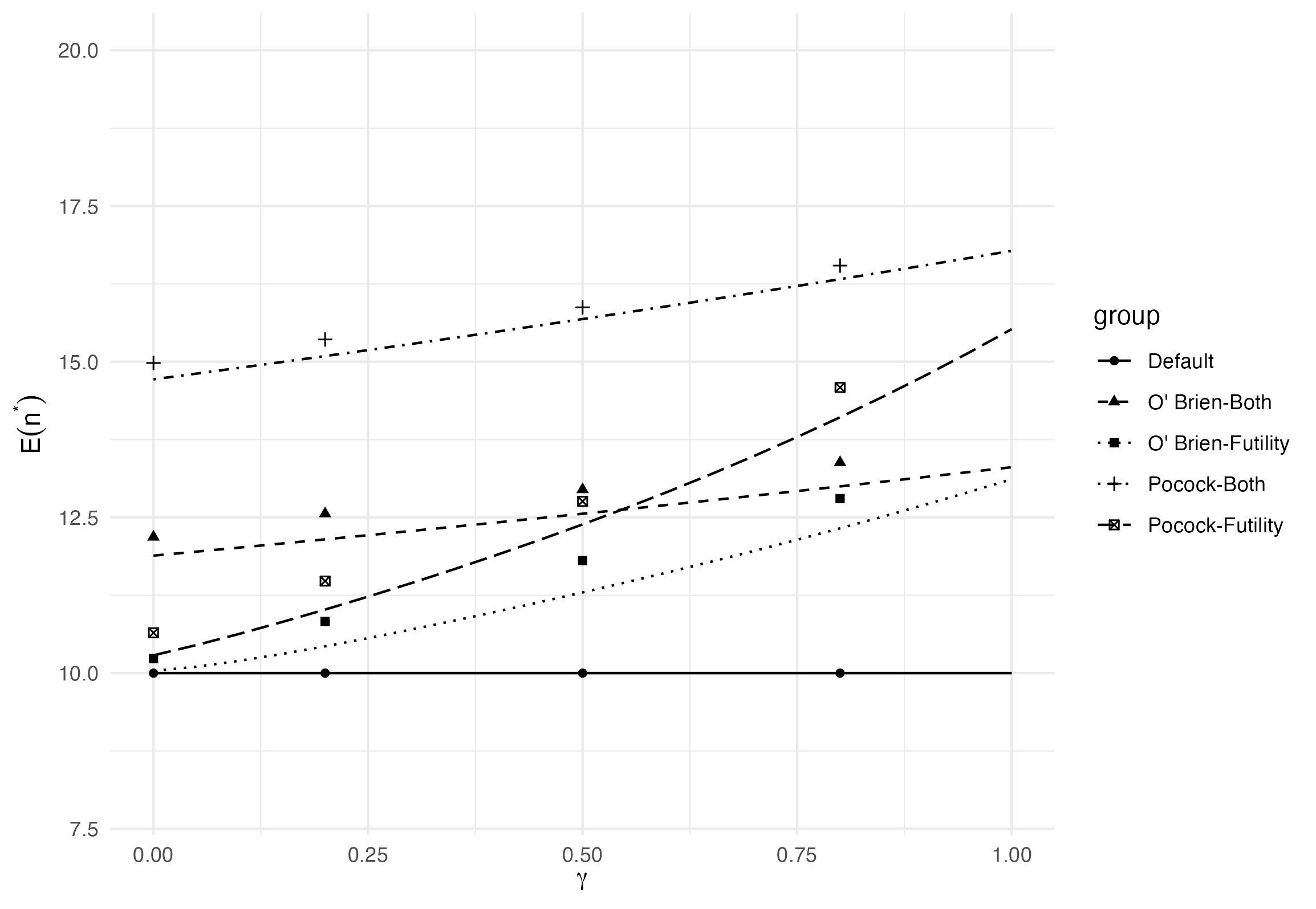}
   \includegraphics[width=6cm , trim=0 0 4cm 0, clip]{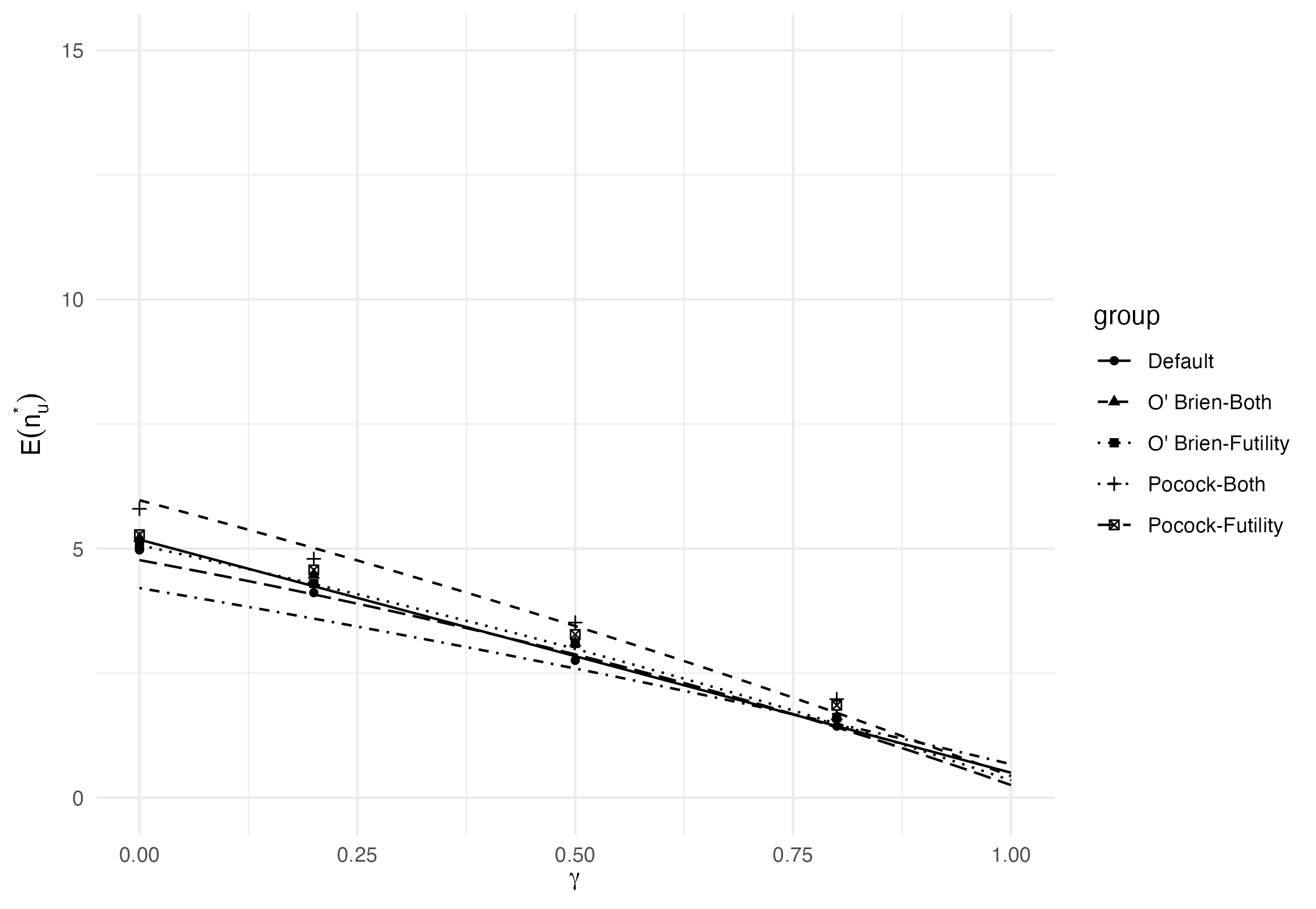}
\includegraphics[width=7.5cm]{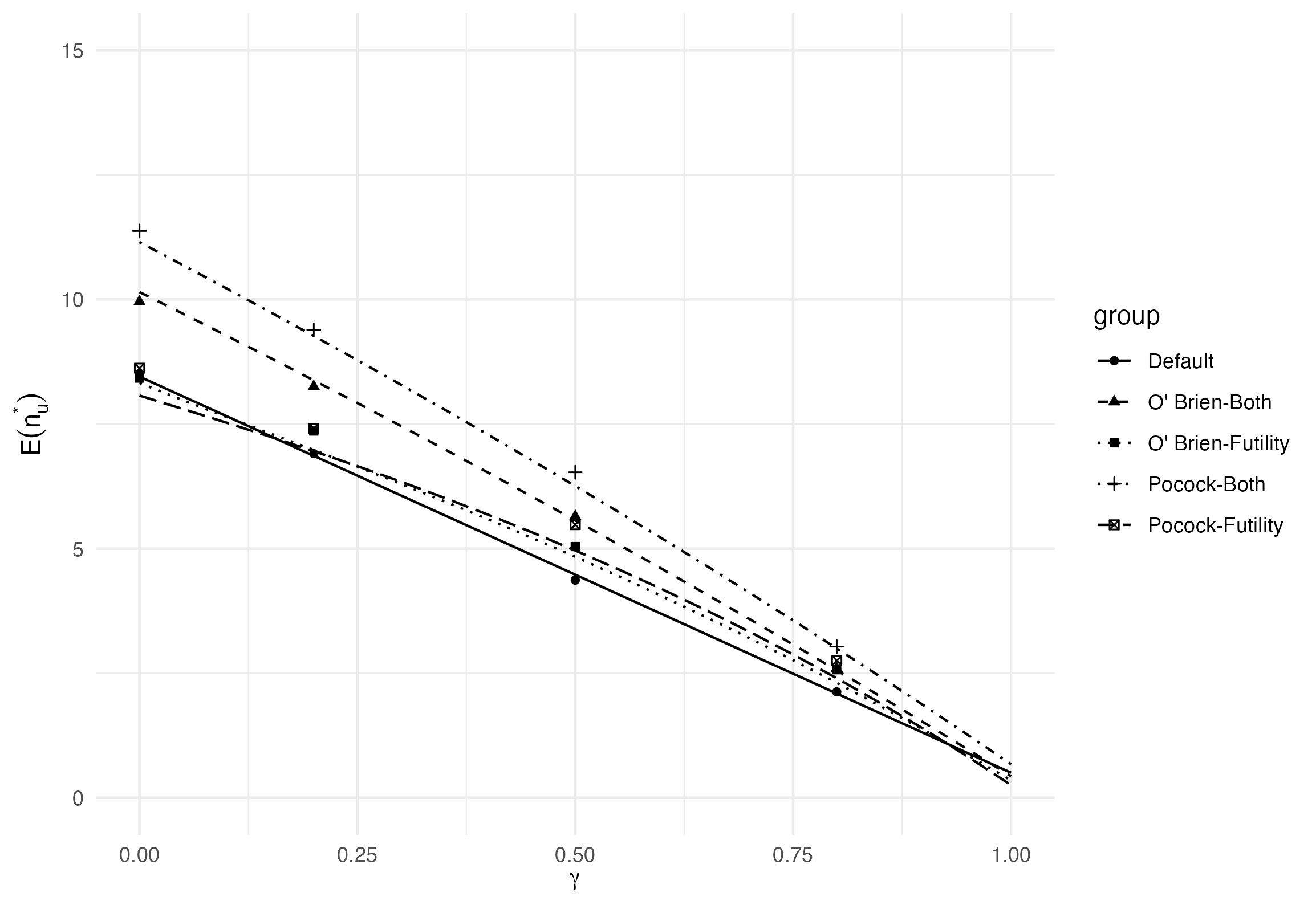}
 \includegraphics[width=6cm , trim=0 0 4cm 0, clip]{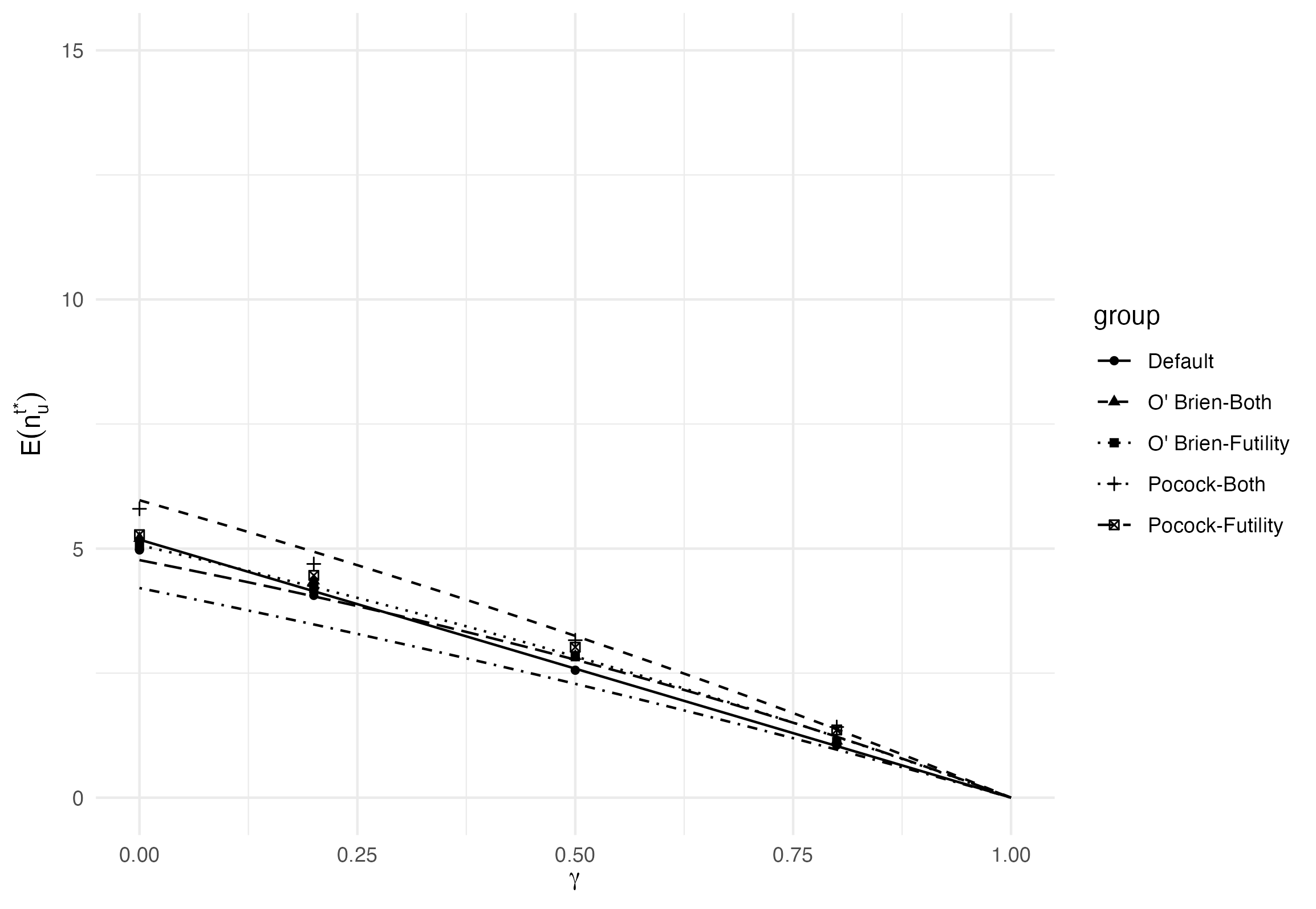}
     \includegraphics[width=7.5cm]{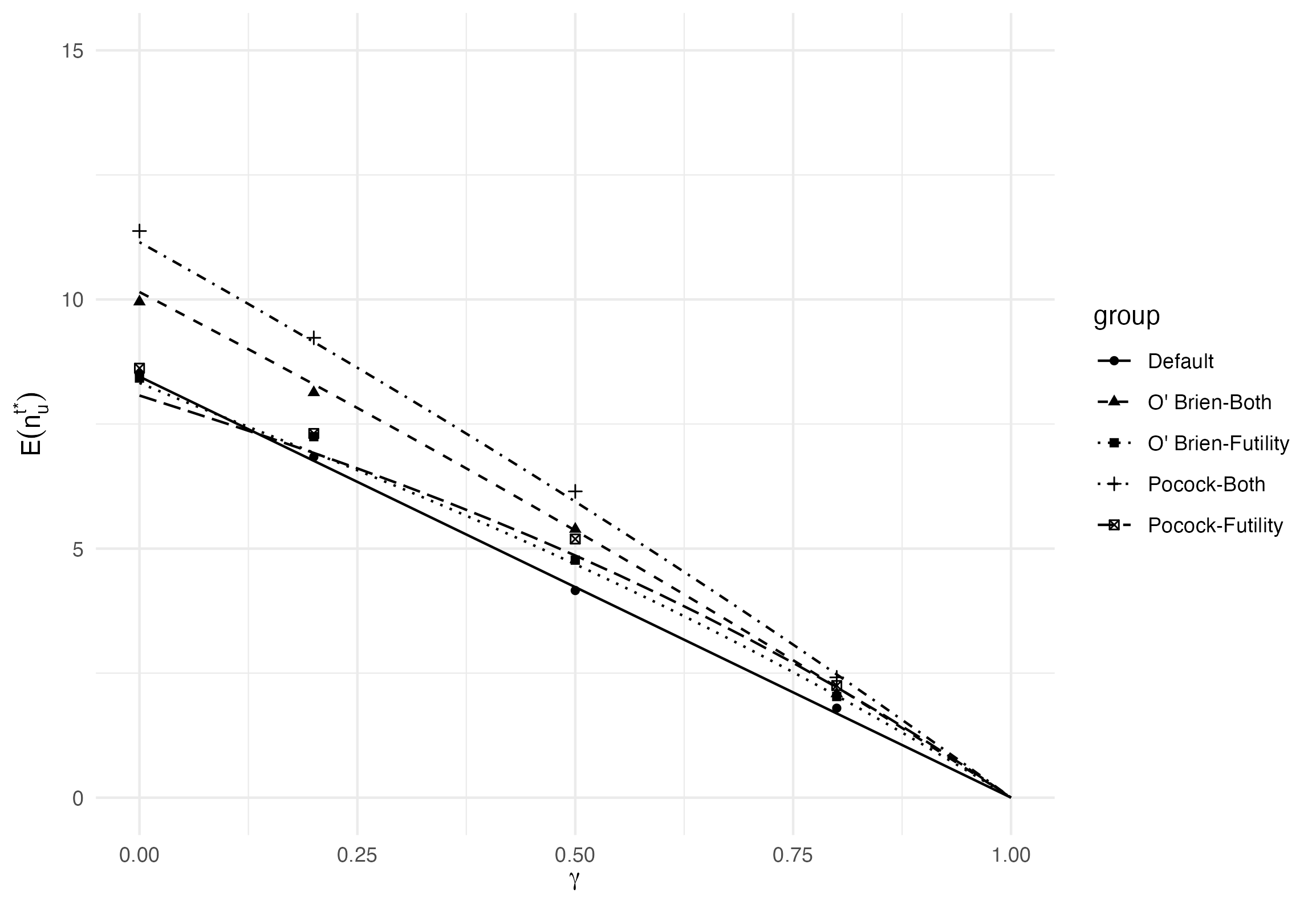}
    \caption{Comparison of the expected number of biomarkers evaluated computed based on the analytical formula and simulation for $t=0.2$ for the misspecified model. We consider two boundary criteria: O' Brien Fleming and Pocock and two stopping criteria: Stopping for both Futility and Efficacy and Stopping only for Futility. Results across combinations of these boundary and stopping criteria are compared with results from a non-sequential procedure, denoted as `Default'. The lines denote the values obtained from the analytical formula, while the dots in several shapes denote the simulation results at $\gamma=0, 0.2, 0.5, 0.8$. The figures to the left correspond to 200 cases and 200 controls, while the figures to the right correspond to 500 cases and 500 controls. }
    \label{fig:rotation_1_0.2}
\end{figure}

\clearpage
\subsection{Probabilities of rejection for a single biomarker panel under a misspecified model}

In this section, we conduct a small simulation study to illustrate the performance of the hypothesis testing for a single biomarker panel, with the test statistic derived based on Theorem 1. 
We begin by generating case-control data with 200 cases and 200 controls for two biomarkers $X_1$ and $X_2$ as follows: \begin{equation*}
   X^{f}|D=0 \sim MVN(0, \Sigma_{{D_0f,sim}})
\end{equation*} and \begin{equation*}
    X^{f}|D=1 \sim MVN( \mu_{f}, \Sigma_{D_1f,sim})
\end{equation*} where $X^{f}=(X_1,X_2)$, $\mu_{f}=(\mu_1, \mu
_2)$, and $\Sigma_{D_1f,sim}$ and $\Sigma_{D_0f,sim}$ are covariance matrices for the cases and controls in which each marker has a variance of 1, and the covariances between the two biomarkers are 0.2 and 0.1, respectively. Thus, we have a misspecified model. We test the following hypothesis at a $5\%$ significance level: \begin{equation*}
    H_0: ROC_f(0.1) \leq 0.56 \ \ \text{vs} \ \  H_0: ROC_f(0.1) > 0.56
\end{equation*} where $\mu_f=(1,1.1)$ corresponds to the null $ROC_f(0.1)=0.56$ computed using 5000 MC samples. The results based on this simulation study are presented in Table 6.

\begin{table}[!htbp]
  \caption{Probabilities of rejection in Stage 1, Stage 2, combined probabilities of rejection, and the probabilities of continuing to Stage 2 for two boundary choices: O'Brien Fleming \& Pocock, with 200 cases, 200 controls for $t=0.1, \lambda= 1/2$.}
    \centering
    \resizebox{\textwidth}{!}{
        \begin{tabular}{cccccccc}
            \hline
            Boundary & Case & $\mu_{f}$ & $ROC(0.1)$ & $S1: P(Reject|H)$ & $S1: P(Continue|H)$ & $S2: P(Reject|H)$ & $P(Reject|H)$ \\ \midrule
          \multicolumn{8}{c}{Both} \\ \hline
            \multirow{3}{*}{O'Brien Fleming} 
            & Null & $(1,1.1)$ & 0.56 & 0.026 (0.005) & 0.490 (0.016) & 0.032 (0.006) & 0.058 (0.007) \\ 
            & \multirow{2}{*}{Alternate} 
            & (1,1.5) & 0.67 & 0.303 (0.015) & 0.627 (0.015) & 0.372 (0.015) & 0.675 (0.015) \\ 
            &  & (0.8, 2) & 0.79 & 0.828 (0.012) & 0.170 (0.012) & 0.170 (0.012) & 0.998 (0.001) \\ \hline
            
            \multirow{3}{*}{Pocock} 
            & Null & $(1,1.1)$ & 0.56 & 0.042 (0.008) & 0.188 (0.012) & 0.017 (0.004) & 0.059 (0.009) \\ 
            & \multirow{2}{*}{Alternate} 
            & (1,1.5) & 0.67 & 0.475 (0.016) & 0.328 (0.015) & 0.186 (0.012) & 0.661 (0.015) \\ 
            &  & (0.8,2) & 0.79 & 0.915 (0.009) & 0.078 (0.008) & 0.072 (0.008) & 0.987 (0.004) \\ \hline

               \multicolumn{8}{c}{Futility} \\ \hline
            \multirow{3}{*}{O'Brien Fleming} 
            & Null & $(1,1.1)$ & 0.56 & 0.000 (0.000) & 0.555 (0.016) & 0.055 (0.008) & 0.055 (0.008) \\ 
            & \multirow{2}{*}{Alternate} 
            & (1,1.5) & 0.67 & 0.000 (0.000) & 0.942 (0.007) & 0.713 (0.014) & 0.713 (0.014) \\ 
            &  & (0.8, 2) & 0.79 & 0.000 (0.000) & 1.000 (0.000) & 0.997 (0.002) & 0.997 (0.002) \\ \hline
            
            \multirow{3}{*}{Pocock} 
            & Null & $(1,1.1)$ & 0.56 & 0.000 (0.000) & 0.367 (0.015) & 0.076 (0.008) & 0.060 (0.008) \\ 
            & \multirow{2}{*}{Alternate} 
            & (1,1.5) & 0.67 & 0.000 (0.000) & 0.840 (0.012) & 0.678 (0.015) & 0.678 (0.015) \\ 
            &  & (0.8,2) & 0.79 & 0.000 (0.000) & 0.996 (0.002) & 0.996 (0.002) & 0.996 (0.002) \\ \hline
        \end{tabular}
    }
    \label{tab:prob_0.1_single}
\end{table}

\end{document}